\documentstyle[11pt,aaspp4]{article}
\def\Msun{~M_\odot}
\def\lsim{\raise0.3ex\hbox{$<$}\kern-0.75em{\lower0.65ex\hbox{$\sim$}}}
\def\gsim{\raise0.3ex\hbox{$>$}\kern-0.75em{\lower0.65ex\hbox{$\sim$}}}

\def\kms{\rm ~km~s^{-1}}
\def\ergs{\rm ~erg~s^{-1}}
\def\ml{~\Msun ~\rm yr^{-1}}
\def\mll{\Msun ~\rm yr^{-1}}

\begin{document}

\title{SUPERNOVA REMNANTS IN MOLECULAR CLOUDS}
\author{Roger A. Chevalier}
\affil{Department of Astronomy, University of Virginia, P.O. Box 3818}
\affil{Charlottesville, VA 22903; rac5x@virginia.edu}

\begin{center}
%version: \today
\end{center}

\begin{abstract}
Massive  ($\gsim 8\Msun$) stars may end their lives in the molecular
clouds in which they were born.
O-type stars probably have sufficient photoionizing radiation and
wind power to clear a region $>15$ pc in radius of molecular material.
Early B stars (B1--B3 on the main sequence, or $8-12\Msun$ stars) 
are not capable of this and may
interact directly with molecular gas.
Molecular clouds are known to be clumpy, with dense molecular clumps
occupying only a few percent of the volume.
A supernova remnant then evolves primarily in the interclump medium,
which has a density $n_H=5-25$ H atoms cm$^{-3}$.
The remnant becomes radiative at a radius of $\sim 6$ pc, forming
a shell that is magnetically supported.
The structure of the shell can be described by a self-similar solution.
When this shell interacts with the dense clumps, the molecular shock
fronts are driven by a considerable overpressure compared to the
pressure in the rest of the remnant.
The expected range of clump sizes leads to a complex velocity distribution,
with the possibility of molecular gas accelerated to a high velocity.
Observations of the remnants W44 and IC 443 can be understood in this model.
W44 has a shell expanding at $\sim 150\kms$ expanding into a medium with
density $4-5$ cm$^{-3}$.  
The shock emission expected in such a model is consistent with the observed
H$\alpha$ surface brightness and the [OI] 63$\mu$m line luminosity.
The clump interaction is seen in OH maser emission, which shows a magnetic
field strength that is consistent with that expected in the model.
IC 443 appears to be expanding at a lower velocity, $100\kms$, into 
an interclump medium with a higher density, $\sim 15$ cm$^{-3}$.
The interaction of the radiative shell with molecular clumps can
produce the  molecular emission that is observed from IC 443.
Both remnants are shell sources of radio synchrotron emission,
which can be attributed to relativistic electrons in the cool radiative shell.
If ambient cosmic ray electrons are further accelerated by the shock
front and by the postshock compression, the radio fluxes and the flat
spectral indices of W44 and IC 443 can be explained.
The energetic electrons are in a high density shell and 
their bresstrahlung emission can approximately
produce the $\gamma$-ray fluxes observed by {\it EGRET}.
Molecular clouds have a significant uniform magnetic field component so
that heat conduction is likely to be important in the hot interior and
can explain the isothermal X-ray emission observed from the remnants.
%Evaporation of the dense clumps by conduction is unlikely to be
%a dominant process.

\end{abstract}


\section{INTRODUCTION}

Twenty years ago DeNoyer (1979a,b) found clear evidence for the
interaction of the supernova remnant IC 443 with molecular gas,
based on observations of disturbed CO emission.
IC 443 has become an especially rich target for
molecular line observations, including H$_2$ (Burton et al. 1988; Richter et al. 1995b);
CO and HCO$^+$ (Dickman et al. 1992; Tauber et al. 1994);
HCN, CN, SiO, CS, SO, H$_2$CO, and C$_2$H (Turner et al. 1992;
van Dishoeck et al. 1993).
Weaker evidence for interaction was found in CO observations of
 W44 and W28 (Wootten 1977, 1981), which were earlier found to be sources
of OH 1720 MHz line emission (Goss \& Robinson 1968).
OH 1720 MHz maser line emission has emerged as a signpost of
molecular interaction with the detection of compact OH emission
from 8 supernova remnants (Frail et al. 1996), including W44, 
IC 443, and 3C391.
These remnants typically have a radius $\sim 10$ pc.

The interaction of supernova remnants with molecular clouds can
plausibly occur because massive stars are born in such clouds and
have a relatively short lifetime.
Stars with initial mass $\gsim 8\Msun$ are expected to end their lives
as core collapse supernovae.
Previous theoretical treatments of the interaction problem
have concentrated on the interaction with
very dense cloud material.
Shull (1980) considered an ambient medium with density $n_o=10^5$ cm$^{-3}$
and predicted strong  IR (infrared) emission that decayed after 20 years.
Wheeler, Mazurek, \& Sivaramakrishnan (1980) found that evolution
in a medium with $n_o=10^4$ cm$^{-3}$ leads to a high IR luminosity
for $\sim 100$ years, followed by a rapid drop.
The remnant radius is $\sim 1$ pc during its bright phase.
Events of this kind have not been identified, but it is not clear
what limits can be set from existing IR observations.
The existing observations of remnant/cloud interactions require
a different model.

The model I explore here is the interaction of a remnant with
a clumpy molecular cloud.
Observations of molecular clouds have shown them to be clumpy,
with a filling factor for the dense molecular clumps of
$\sim 0.02-0.08$ (Blitz 1993).
The interclump gas has a density of $5-25$ cm$^{-3}$.
A supernova is more likely to explode in the interclump gas than
in a dense clump, especially when the effects of presupernova winds
and photoionization and stellar motion are taken into account.
In \S~2, I consider the likely environment for a supernova in
a molecular cloud, including the effects of presupernova winds
and photoionization as well as the structure of the molecular cloud.
The remnant expansion in a molecular cloud is treated in \S~3,
on the assumption that the evolution in the interclump region 
can be separated from the interaction with clumps.
In \S~4, I compare this model to observations of three well-observed
remnants: W44, IC 443, and 3C391.
The model does appear to be compatible with the basic observations.
Nonthermal particles and their emission are treated in \S~5.
This topic is of special interest in view of the probable detection
of supernova remnants in molecular clouds with the {\it EGRET}
experiment on the {\it CGRO} ({\it Compton Gamma-Ray Observatory})
(Esposito et al. 1996).
Possible future work is briefly discussed in \S~6.

\section{THE SUPERNOVA ENVIRONMENT}

Observational studies of molecular clouds have shown them
to have dense molecular clumps embedded in a lower density interclump
medium (Blitz 1993).
A good example is the study of the Rosette cloud by Williams,
Blitz, \& Stark (1995),
who show that the clumps have a density of 440 H atoms cm$^{-3}$
and a filling factor of 0.08.
The interclump medium has a density of
11 H atoms cm$^{-3}$, so that
the clumps comprise 3.5 times more mass than the interclump medium, but
occupy a small fraction of the volume.
The distribution of clump masses, $M$, in clouds can be described by a power law,
$dN(M)\propto M^{-1.54}dM$, where $dN$ is the number of clouds with
mass between $M$ and $M+dM$.
The relevant mass range is $1-3,000\Msun$, which corresponds to diameters
of $0.2-2.2$ pc if clumps have a typical H$_2$ density of $10^3$ cm$^{-3}$.
More generally, Blitz (1993) notes some properties of typical GMC's
(Giant Molecular Clouds):  total mass $\sim 10^5\Msun$, diameter
$\sim 45$ pc,  H$_2$ density in clumps of $10^3$ cm$^{-3}$,
 interclump gas  density in the range $5-25$ H atoms cm$^{-3}$,
and clump filling factor of $2-8$ \%.
The clump mass spectrum can be described by $dN(M)\propto M^{-q}dM$ with
$q$ in the range $1.4-1.9$.
More recently, Kramer  et al. (1998) have also found power law distributions
of clump mass with $q\approx 1.6-1.8$ in 7 clouds;  the clump masses
range from $10^{-4}-10^3\Msun$.
With this mass spectrum, most of the clouds are small, but most of
the mass is in large clumps.
The lower limit to the clump masses is set by observational constaints.
Blitz (1993) noted that the clump shapes vary from spherical to
highly irregular.

An interclump pressure, $p/k\approx 10^5$ K cm$^{-3}$, is needed to
confine the dense clumps and to support the cloud against gravitational
collapse (Blitz 1993).
The thermal pressure in the interclump gas is clearly too small and
magnetic fields are a plausible pressure source.
Studies of the polarization of the light from stars behind molecular
clouds implies that the interclump magnetic field does have
a relatively smooth structure (Heiles et al. 1993).
Setting $B_o^2/8\pi$ equal to the above pressure yields $B_o=19p_{5}^{1/2}~\mu$G,
where $B_o$ is the uniform field component and $p_5$ is $p/k$ in
units of $10^5$ K cm$^{-3}$.
However, a uniform magnetic field would not provide support along the
magnetic field and would not explain the large line widths observed
in clouds.
An analysis of the polarization and Zeeman effect in the dark cloud
L204 shows consistency with equal contributions to the pressure
from a uniform field and a fluctuating, nonuniform component
(Myers \& Goodman 1991; Heiles et al. 1993).
The uniform component is then $B_o=13p_{5}^{1/2}~\mu$G.
The magnetic field strength deduced in L204 is consistent with
this value.

The molecular cloud can be influenced by stellar mass loss and
photoionization by the progenitor star before the supernova.
The effect of these processes can be to remove molecular gas from
the vicinity of the central star.
From the point of view of interpreting the observations of supernova
remnants mentioned in \S~1, it is important whether molecular clumps
can survive within $\sim 15$ pc of the central star.
McKee, Van Buren, \& Lazareff (1984) found that B0--O4 stars
could clear a region 28 pc in radius of clumps if the mean density
was $n_m=10$ cm$^{-3}$.
Small clumps are completely photoevaporated and large clumps
move out by the ``rocket'' effect.
If the scaling for the radius $\propto n_m^{-0.3}$ 
(McKee et al. 1984) can be extrapolated
to the somewhat higher mean densities of interest here,
a region $\gsim 15$ pc is expected to become free of molecular clumps.
The photoevaporation of clumps has been studied in greater
detail by Bertoldi \& McKee (1990), who found that the rocket
effect is smaller than that in previous investigations.
However, the rocket velocity $\gsim 5\kms$ is still sufficient to
clear a region $\gsim 15$ pc in size.
The additional effect of stellar winds is to clear a wind bubble in
the homogenized region for the stronger winds or to create a wind
bubble within the HII region (McKee et al. 1984).

The situation changes for early B type stars, which have weaker
photoionizing fluxes and stellar winds.
These stars have sufficiently long lifetimes that their HII regions
can come into pressure equilibrium with the surrounding medium.
I assume that the surrounding interclump medium exerts a pressure
$p/k=10^5$ K cm$^{-3}$ and that the gas temperature in the HII
region is $10^4$ K, so that the density in the ionized gas
is $\sim 5$ H atoms cm$^{-3}$.
For the ionizing fluxes given in Panagia (1973), Table 1 shows
the radii of the ionized regions, $R_{\rm ionized}$, for O9--B3 main sequence stars.
The small sizes of the ionized regions for B1--B3 stars is evident.
The radii for the regions around the O9 and B0 stars appears to
be smaller than those obtained by McKee et al. (1984); this is
presumably because of the higher surrounding pressure taken here.

Stellar winds are an additional effect on the circumstellar environment.
The maximum size, $R_b$, of the wind bubble is that at which the bubble
comes into pressure equilibrium with its surroundings, or the wind energy
production is about equal to the displaced energy:
\begin{equation}
{1\over 2} \dot M v_w^2 \tau_{ms} = \left({4\over 3}\pi R_b^3\right)
 {3\over 2} p_o,
\end{equation}
where $\dot M$ and $v_w$ are the wind mass loss rate and velocity and
$\tau_{ms}$ is the main sequence age.
For typical parameters, the result for $R_b$ is
\begin{equation}
R_b=5.8\left(\dot M\over 10^{-8} \ml\right)^{1/3}
\left(v_w\over 700\kms\right)^{2/3}
\left(p/k\over 10^5 {\rm~K~cm^{-3}}\right)^{-1/3}
\left(\tau_{ms}\over 10^7 {\rm~yr}\right)^{1/3}  {\rm~pc}.
\end{equation}
The values of $\dot M$ in Table 1 are from de Jager,
Nieuwenhuijzen,  \& van der Hucht (1988).
The values of $\tau_{ms}$ are estimated from the evolutionary
tracks of Schaller et al. (1992) for a somewhat different set
of stellar masses.
I have taken $v_w=700\kms$, as appropriate for these
 stellar types, and $p_o/k=10^5$ K cm$^{-3}$ in
order to obtain the values of $R_b$ shown in Table 1.
It can be seen that the wind bubble is somewhat larger than the
ionized region, although B1--B3 stars do have bubble sizes that
are smaller than those of the remnants of interest.
McKee et al. (1984) note that if the wind bubble begins to expand
beyond the ionized region, matter will be photoevaporated from
clumps and may mix with the hot gas inside the bubble.
The higher density gas can have a cooling time that is shorter than
the age, so that the assumption of energy conservation of the bubble gas
made in eq. (1) is not valid, and the bubble radius may be limited
to a value close to that of the ionized region.

In the later stages of evolution, the star becomes a red
supergiant and has a slow dense wind which can cool after passing
through a termination shock.
Equating the wind ram pressure ($\rho_w v_w^2$) with the interclump
pressure, $p_o/k=10^5$ K cm$^{-3}$, yields an approximate stopping radius of
\begin{equation}
r_{st}=0.62 \dot M^{1/2}_{-5}v_{w1}^{1/2}\quad {\rm pc},
\end{equation}
where $\dot M_{-5}$ is the wind mass loss rate in units of
$10^{-5}\ml$ and $v_{w1}$ is the wind velocity in units of $10\kms$.

Even if the immediate progenitor star does not modify its environment,
it is quite possible that  more massive stars form in association with
the progenitor and that these stars affect the environment.
The $8-12\Msun$ stars explode relatively late so that more massive
companions do have an opportunity to change the environment.
However, the lower mass molecular clouds (mass $\sim 10^4\Msun$) can
have one or no O stars (Williams \& McKee 1997).
This discussion leads to the conclusion supernova remnants with
radii $\sim 10$ pc can involve interaction with a quiescent
molecular cloud, although they are a  fraction of the Type II supernovae.
The examination of remnants where cloud interaction appears to be
taking place will then indicate whether this is occurring, or whether
modification of the supernova surroundings by the progenitor
star and its companions needs to be taken into account.

\section{SUPERNOVA EXPANSION IN A MOLECULAR CLOUD}

\subsection{Expansion in the Interclump Medium}

A reference interclump density is $n_o=10$ H atoms cm$^{-3}$.
I assume that as a first approximation, the presence of the dense
clumps can be neglected in the propagation of the supernova
shock front.
The evolution of a supernova remnant in a uniform medium has
been frequently treated, so I primarily highlight some of the
aspects that are important for the molecular cloud case.

In the first phase of evolution, the supernova ejecta are decelerated by
interaction with the ambient medium.
The ejecta are expected to be mostly freely expanding until they
have swept up their own mass, which occurs at a radius 
$R_m=1.9 M_{1}^{1/3}n_1^{-1/3}$ pc, where $M_{1}$ is the ejected mass
in units of $10\Msun$ and $n_1$ is the interclump H density in units
of 10 H atoms cm$^{-3}$.
This radius is sufficiently small that the effects of the progenitor
star on its surroundings are likely to be important for this initial
phase of evolution.
However, once the shock front has propagated several pc, the
flow in the interclump medium should tend toward the Sedov blast wave solution.
%The ejecta steadily mix with the interclump gas because of
%Rayleigh-Taylor instabilities (Chevalier, Blondin, \& Emmering 1992),
%but once the radius has passed $R_m$, the mixing becomes weak.

The remnant next enters a radiative phase (Cox 1972; Chevalier 1974;
Cioffi, McKee, \& Bertschinger 1988).
For gas with solar metallicity, Cioffi et al. (1988) find that the
time taken to cool to 0 temperature, i.e. shell formation, is 
$t_{sf}=9.7\times 10^3 E_{51}^{3/14} n_1^{-4/7}$ years, where $E_{51}$
is the supernova energy in units of $10^{51}$ ergs.
The effects of radiative losses actually become important at an
earlier time and Cioffi et al. suggest that the pressure-driven snowplow
(PDS) phase begins at at $t_{\rm PDS}=t_{sf}/e=3.6\times 10^3
 E_{51}^{3/14} n_1^{-4/7}$ years, where $e$ is the base of the natural
logarithm.
The corresponding shock radius and velocity are $R_{\rm PDS}=
5.2 E_{51}^{2/7} n_1^{-3/7}$ pc and $v_{\rm PDS}=574 E_{51}^{1/14} n_1^{1/7}\kms$.
For the further evolution of the remnant radius, Cioffi et al. (1988)
advocate an offset power law in time.
However, over the range of $(5-50)t_{\rm PDS}$ the value of $v_s t/R$ is
in the range $0.31-0.33$ so that a power law in time is an adequate representation.

For $p/k=10^5$ cm$^{-3}$ K and $n_o=10$ cm$^{-3}$,
the isothermal sound speed in the cloud is $7.7\kms$.
For $E_{51}=1$ and $n_1=1$, 
the shock velocity drops to twice the isothermal sound speed
at a radius of 25 pc.
Because the typical diameter of a cloud is 45 pc (Blitz 1993),
the remnant is more likely to break out of the cloud  before it
merges with the cloud.

\subsection{Radiative Shell Structure}

The expansion of the radiative shell can be treated independently
of the shell structure, which depends on the radiative cooling and
on contributions to the pressure.
This is true provided that the shell thickness is small compared
to its radius ($\Delta R/R\ll 1$).
Bertschinger (1986) developed an analytical solution for the shell
structure on the assumption of radiative cooling to a particular temperature
and adiabatic evolution after that point; magnetic effects were neglected.
Cioffi et al. (1988) made the same assumption in their numerical
calculations, allowing radiative cooling to $1.2\times 10^4$ K.
On the other hand, Chevalier (1974) and Slavin \& Cox (1992) performed numerical calculations
with  magnetic support in the dense shell.

The observations of molecular clouds point to the presence of
a significant uniform magnetic field.
I take the case of expansion in a uniform magnetic field and start
with the expansion perpendicular to the magnetic  field direction.
If a uniform magnetic field gives a pressure of $p/k=10^5$ cm$^{-3}$ K,
the field strength is $1.9\times 10^{-5}$ G.
The field is likely to have a significant nonuniform component,
but a tangential field component $\sim 10^{-5}$ G is plausible.
The density at the point where the magnetic pressure becomes important
can be expressed as $n_m=2.4\times 10^2 v_2 n_1^{3/2} B_{ot-5}^{-1}$ cm$^{-3}$,
where $v_2$ is the shock velocity in units of 100 $\kms$ and
$B_{ot-5}$ is the ambient tangential field in units
of $10^{-5}$ G (e.g., Draine  \& McKee 1993).
The immediate postshock temperature is $1.38\times 10^5 v_2^2$ K
if the gas is fully ionized, so that
the tangential field comes to dominate the pressure provided that
the temperature drops below $3\times 10^4 v_2 n_1^{-1/2} B_{ot-5}$ K;
the radiative cooling is certainly likely to take the temperature
below this value and the cooling becomes more isochoric than isobaric
as in the case of a lower magnetic field. 
The magnetic field strength after the compression is
$B=2.4\times 10^{-4} n_1^{1/2} v_2$ G and the Alfven velocity is
substantial, $28 v_2^{1/2} n_1^{-1/4} B_{ot-5}^{1/2} \kms$.
If we consider a ring of radius $R$ and thickness $\Delta R$
perpendicular to the field direction, flux conservation implies that
the quantity $B_t2\pi R \Delta R$ is conserved.
Mass conservation implies $\rho 4\pi R^2\Delta R$ is conserved, for
spherical expansion.
The result is  $B_t\propto \rho R$.
This is an approximation because the expansion is not exactly spherical.

The method of Bertschinger (1986) for the shell structure can
be extended to the magnetic case of interest here.
The equations of mass and momentum conservation for spherical flow are
\begin{equation}
{d\rho\over dt}+{\rho\over r^2}{\partial r^2 v\over\partial r}=0, \quad
\rho{dv\over dt}+{1\over 8\pi}{dB_t^2\over dr}+{B_t^2\over 4\pi r}=0,
%{dp\over dt}-{\gamma p\over \rho}{d\rho\over dt}=0,
\end{equation}
where $d/dt=\partial / \partial t+v\partial / \partial r$ and 
 the magnetic pressure is taken to dominate in the dense shell.
Flux conservation can be written as
\begin{equation}
{d(B_t/\rho r)\over dt}=0.
\end{equation}

I assume that the outer shock has a radius $r_s\propto t^{\eta}$,
so that the shock velocity is $v_s\propto t^{\eta-1}$.
As in Bertschinger (1986), the  compression in the radiative,
thermal pressure-dominated region
is described by
a parameter $\lambda_c=\rho_o/\rho_c$, where $\rho_o$ is the ambient
density and $\rho_c$ is the density at the base of the radiative region.
For a case with no magnetic field and an isothermal shock front (as
in Bertschinger 1986), $\lambda_c\propto {\cal M}^{-2}\propto v_s^{-2}$, where
$\cal M$ is the shock Mach number.
In the present case, the compression is limited by the magnetic field
and $\lambda_c\propto {\cal M}^{-1}_a\propto v_s^{-1}$, where
${\cal M}_a$ is the Alfven Mach number.
The thickness of the dense shell is $\sim \lambda_c r_s$, so the
radial variable is changed to $\zeta$, defined by
$r=r_s(1-\lambda_c\zeta)$.
The radiative, thermally supported region is taken to be of negligible
thickness, which should be a good approximation.
Dimensionless variables for the physical variables can be defined by
\begin{equation}
v={r_s\over t}(\eta-\lambda_c W_o), \quad
\rho=\rho_o\lambda_c^{-1}E_o, \quad
{B_t^2\over 8\pi}=\rho_o\left( r_s\over t\right)^2(1-\lambda_c)P_o.
\end{equation}
The boundary conditions at $\zeta=0$ are $W_o=\eta$, $E_o=1$, and $P_o=\eta^2$.

The main assumption made here is that the shell thickness is much
less than its radius, so that $\lambda_c$ can be regarded as a
small parameter.
Substitution of eqs. (6) into eqs. (4) and (5) yields
\begin{equation}
x{d\ln E_o\over d\zeta}+{dW_o\over d\zeta} +3\eta-1=0,
\end{equation}
\begin{equation}
{dP_o\over d\zeta}+\eta(1-\eta)E_o=0,
\end{equation}
\begin{equation}
x\left({d\ln P_o\over d\zeta}-2{d\ln E_o\over d\zeta}\right)-2\eta=0,
\end{equation}
where
\begin{equation}
x=W_o-\zeta
\end{equation}
describes the gas velocity relative to a frame moving with the
shell.
The magnetic tension term does not appear in eq. (8) because it
is higher order in $\lambda_c$.
This is consistent with the numerical results of Slavin \& Cox (1992), who
found that inclusion of this term did not substantially change their results.
Integration of eqs. (7)--(10) can be initiated at $\zeta=0$ and carried
to the inner boundary, where $x=0$.
Results for $\eta=0.3$ are shown in Fig. 1.
In contrast to the non-magnetic, adiabatic case where the density is
constant over much of the shell and becomes large at the inner boundary
(Bertschinger 1986), here the density decreases in from the shock
front and drops at the inner boundary.

In their numerical calculations, Slavin \& Cox (1992) prefer the
relation $d(B_t/\rho)/dt=0$ to that given in eq. (5) in order to
give a lower bound on the magnetic effects.
In this case, the flow in the shell is isentropic with $\gamma=2$
and the constant term drops out of eq. (9).
The equations admit an analytic solution:
\begin{equation}
E_o=1-{1-\eta\over 2\eta}\zeta,\quad
P_o=\eta^2\left(1-{1-\eta\over 2\eta}\zeta\right)^2,
\end{equation}
\begin{equation}
W_o={3\eta^2\over 1-\eta}+\left(1-{3\over 2}\eta\right)\zeta-{2\eta^2(4\eta -1)\over
(1-\eta)[2\eta-(1-\eta)\zeta]}.
\end{equation}
The inner boundary occurs at 
\begin{equation}
\zeta_i={2\eta\over 1-\eta}\left[ 1-\left( 4\eta -1\over 3\eta\right)^{1/2}\right]. 
\end{equation}
Results for this case for $\eta=0.3$ are shown in Fig. 2.
The inner edge of the shell is at $\zeta_i=0.4531$.
The density contrast across the shell is smaller than in the
previous case, which brings the results closer to the non-magnetic
case, as expected.
The results compare well to the numerical computations of Slavin \&
Cox (1992, fig. 1) during the time that the shell thickness is
small compared to the radius.

Because there are approximations made in  describing a 2-dimensional
situation by a 1-dimensional  flow, it is instructive to consider
the 1-dimensional case of cylindrical expansion into a uniform
magnetic field.
The expansion is perpendicular to the field direction.
The mass conservation equation is now
\begin{equation}
{d\rho\over dt}+{\rho\over r}{\partial r v\over\partial r}=0, 
\end{equation}
so that eq. (7) becomes
\begin{equation}
x{d\ln E_o\over d\zeta}+{dW_o\over d\zeta} +2\eta-1=0.
\end{equation}
Now, mass and magnetic flux conservation imply that $B/\rho$ is
conserved exactly.
The flow in the shell is again isentropic so that $E_o$ and $P_o$
are described by eqs. (11).
The velocity variable is given by
\begin{equation}
W_o={2\eta^2\over 1-\eta}+(1-\eta)\zeta-{2\eta^2(3\eta -1)\over
(1-\eta)[2\eta-(1-\eta)\zeta]}.
\end{equation}
and the inner boundary occurs at 
\begin{equation}
\zeta_i={2\eta\over 1-\eta}\left[ 1-\left( 3\eta -1\over 2\eta\right)^{1/2}\right].
\end{equation}
For cylindrical expansion, the case of momentum conserving
expansion is $\eta=1/3$,
so that now $\eta\ge 1/3$.
The solution is very close to the one illustrated in Fig. 2.

\subsection{Interaction with Clumps}

The supernova remnant becomes radiative in the interclump medium
at a relatively small radius ($\sim 5$ pc) and interaction of the cool shell
with a molecular clump becomes a matter of interest.
This is the phase of evolution that is most likely to be observed because
it lasts longer than the early adiabatic phase, and the observations
of remnants discussed in the next section imply that the remnants 
under consideration are
in the radiative phase.
In the initial interaction, the radiative shell drives a dense
slab into the clump, as illustrated in Fig. 3.
Here, I approximate the motion of the slab as one-dimensional.
The slab is bounded by shock waves in the clump and in the radiative shell.
The equations describing the evolution of the position $x$,
surface density $\sigma$, and velocity $v$ of the slab are
\begin{equation}
{dx\over dt}=v,
\end{equation}
\begin{equation}
{d\sigma\over dt}=\rho_{rs} (v_{rs}-v)+\rho_{cl}v,
\end{equation}
\begin{equation}
{dv\over dt}={\rho_{rs} (v_{rs}-v)^2-\rho_{cl}v^2\over\sigma},
\end{equation}
where $\rho_{rs}$ and $v_{rs}$ are the density and velocity in
the radiative shell.
Provided the thickness of the radiative shell is sufficiently small,
$v_{rs}$ can be regarded as constant (see \S~3.2).
However, the density $\rho_{rs}$ can vary with position in the shell.

In eq. (20), I assume that the 2 shock fronts bounding the slab are strong.
Although this is probably an excellent approximation for the shock front
in the molecular clump, it is not such a good approximation for the shock
in the radiative shell because of the high Alfven velocity in the
shell (see \S~3.2).
The shock in the shell may be a weak shock.
I neglect that fact here  in order to estimate the slab motion.

The case of constant density $\rho_{rs}$ is particularly simple.
The slab then moves at a constant velocity
\begin{equation}
v=v_{rs}\left[ 1+\left(\rho_{cl}\over\rho_{rs}\right)^{1/2}\right]^{-1}.
\end{equation}
The ram pressures generated by the two shock fronts are equal to
each other.
The ratio of the ram pressure in the slab to that at the front of the
radiative shell is
\begin{equation}
{\rho_{cl} v^2\over\rho_o v_{rs}^2}=
{\rho_{cl}\over\rho_o}\left[1+\left(\rho_{cl}\over\rho_{rs}\right)^{1/2}\right]^{-2}.
\end{equation}
The effect of the magnetic field is to limit $\rho_{rs}$ to somewhat
below $\rho_{cl}$ so that the pressure ratio may be $\sim 10-100$.
While the shock fronts are in the dense gas, the ratio of the swept up
column density of clump gas to that from the radiative shell is
simply $(\rho_{cl}/\rho_{rs})^{1/2}$.
If the quantity $\sigma_{cl}\rho_{rs}^{1/2}/\sigma_{rs}\rho_{cl}^{1/2}$ is
$< 1$, the shock front breaks out of the clump first, but if it is
$> 1$, the shock front breaks out of the radiative shell first.
Here, $\sigma_{cl}$ and $\sigma_{rs}$ are the total column densities
through the clump and the radiative shell, respectively.
If the shock front breaks out of the clump first, then the shocked
clump is initially accelerated by the continued impact of the radiative shell.

For a more general density structure, it is useful to change to
dimensionless variables.
The position of the slab in the radiative shell is given by
the value of $\zeta$ at the slab $\zeta_s$ and the radiative shell
density can be described by $E_o(\zeta)$ as derived in \S~3.2.
In addition, I define
\begin{equation}
y={t\over \lambda_c\tau},\quad U_o={v\over v_{sh}}, {\rm ~and}
\quad \Omega={\sigma\over \rho_{so}v_{rs}\lambda_c\tau},
\end{equation}
where $\tau=\eta r_s/v_{rs}$ is the age of the remnant.
The substitution of these variables into eqs. (18)-(20) yields
\begin{equation}
{d\zeta_s\over dy}=(1-U_o)\eta,
\end{equation}
\begin{equation}
{d\Omega \over dy}=E_o(\zeta_s)(1-U_o)+\alpha U_o,
\end{equation}
\begin{equation}
{dU_o\over dy}={E_o (1-U_o)^2-\alpha U_o^2\over \Omega}
\end{equation}
where $\alpha \equiv \rho_{cl}/\rho_{so}$.
Results for the case $E_o(\zeta)$ given by eq. (11) are shown in fig. 4.
The slab receives an impulse from the initial dense part of
the radiative shell that carries it forward even when the
driving density in the shell drops.
The slab decelerates, so that the forward pressure is
greater than that at the shock front in the radiative shell,
given by $\sigma_{rs}v_{rs}$, where $\sigma_{rs}$ is the 
mass column density through the radiative shell.

For the larger molecular clumps, the shock front breaks out of the
rear of the radiative shell while the forward shock is in the
molecular clump.
Whatever the density structure in the radiative shell, the
shell imparts a definite momentum to the motion of the slab.
Initially, the interior pressure of the supernova remnant is
small compared to the ram pressure of the slab, and the
evolution of its position, $x$, can be described by
\begin{equation}
(\sigma_o+x\rho_{cl}){dx\over dt}=\sigma_{rs}v_{rs}.
\end{equation}
where $\sigma_o$ is the column density in the slab at the end of
the double shock phase.
Once the swept-up column density becomes $\gg \sigma_o$, the
position of the slab is
\begin{equation}
x\approx\left(2\sigma_{rs}v_{rs}t\over \rho_{cl}\right)^{1/2}
\end{equation}
which describes the ``snowplow'' evolution in one  dimension.
The time $t$ here is initiated at the end of the double shock phase.
If the remnant is well into the radiative phase, $\sigma_{rs}\approx
\rho_o R/3$.
The time it takes the slab to traverse a distance $x$ can be written as
\begin{equation}
{t\over\tau}={3\over 2\eta}\left(\rho_{cl}\over\rho_o\right)
\left(x\over R\right)^2,
\end{equation}
where $\tau$ is the age of the remnant and $t<\tau$ has been assumed
in deriving this expression.
For the range of clump sizes expected in a molecular cloud, the
small ones can be crushed soon after being hit by the radiative shell
($t<\tau$), while the large ones continue to be shocked after
they are left inside the remnant.

The approximate expression for the slab position (eq. [28]) can be used
to find the surface density: $\sigma_o=x\rho_{cl}\approx 
(2\sigma_{rs}v_{rs} \rho_{cl}t)^{1/2}$.
The variables in eq. (23) can then be used to find an expression for
$\Omega$ that should be approached at late times: $\Omega=(2 \alpha y/
3\eta)^{1/2}$, where $\alpha \equiv \rho_{cl}/\rho_{so}$.
Fig. 4 shows that this provides a good representation of the column density.
The evolution derived here assumes that the slab is entirely driven by
the impulse from the radiative shell and the pressure of the hot interior
gas can be neglected.
This assumption can be justified.
The ram pressure in the slab is initially high and, according to
eq. (28), is later equal to $\sigma_{rs}v_{rs}/t$.
After a time  about equal to the initial age, $\tau\approx 0.3 R/v_{rs}$,
we have the ram pressure $\approx \rho_o v_{rs}^2$.
The hot interior pressure $\lsim 0.3 \rho_o v_{rs}^2$ (Chevalier 1974;
Cioffi et al. 1988).
At later times, the ram pressure evolves as $t^{-1}$, where $t$ now
approaches the time since the explosion.
If the hot interior pressure evolves adiabatically, it varies as
$R^{-5}\propto t^{-5\eta}\propto t^{-1.5}$ and thus drops more rapidly
than the ram pressure.
The hot interior pressure is most important after a time about equal to
the initial age, but even then it is not very significant.
Radiative losses of the hot gas can further reduce the pressure.
This situation can be contrasted with that for a cloud in an adiabatic
blast wave, for which the evolution of the interior pressure is
 important (McKee et al. 1987).

\subsection{The Relativistic Particle Component}

Consideration of relativistic electrons and protons in the
supernova remnant is important for radio synchrotron radiation
and for high-energy $\gamma$-ray emission.
I will assume that the relativistic particles in the dense shell are produced
in the shock front, although a central pulsar is another potential
source of relativistic particles.
I examine the case of propagation perpendicular to the magnetic
field and assume that the relativistic particles are adiabatically
compressed and maintain isotropic pitch angles.
The gyroradii of the particles are
 smaller than the cooling length and the thickness of
the cool shell, so that the adiabatic approximation is adequate.

There is not a well-developed theory for the shock acceleration of
relativistic particles in a evolved remnant, so the considerations
here are guided by the observations.
In Blandford \& Cowie (1982) and in \S~5 here, evidence is presented
that the immediate postshock energy in relativistic particles is a small
fraction of the total postshock energy density and that the acceleration of
Galactic cosmic rays is an adequate source of relativistic particles.
In this case, a constant fraction of the flux of particles
entering the shock front is injected into the relativistic
particle population.
The pressure of relativistic particles in the immediate
postshock region is then $p_{rs}=\beta \rho_o$.
The postshock evolution is adiabatic with $\gamma_r=4/3$.
Assuming that the relativistic particle pressure does not become
important in the cooling region, the relativistic pressure at
the base of the radiative cooling region is
$p_{rc}=p_{rs}/(4\lambda_c)^{4/3}$.

The further evolution of the relativistic particle pressure, $p_r$, is
then described by
\begin{equation}
{dp_r\over dt}-{4 p_r\over 3\rho}{d\rho\over dt}=0.
\end{equation}
The solution for $p_r$ in the cool shell can be found from the
solution for the shell in \S~3.2 provided that $p_r$ does not
become an important contributor to the total pressure.
The dimensionless variable for the relativistic pressure is
defined by $p_r=p_{rc}P_r$ and eq. (30) becomes
\begin{equation}
{d\ln P_r\over d\zeta}-{4\over 3}{d\ln E_o\over d\zeta}=0.
\end{equation}
The relativistic fluid in the
shell is isentropic so that $P_r=E_o^{4/3}$.
The case of $B_t\propto \rho$ (see eq. [11]) gives
$P_r=[1-(1-\eta)\zeta/2\eta]^{4/3}$.
If the relativistic pressure does become a significant part of the
total pressure, the relativistic pressure must be included in the
momentum equation and the dense shell becomes wider.

\section{PROPERTIES OF OBSERVED REMNANTS}

A number of remnants are likely candidates for interaction with
molecular clouds.
Frail et al. (1996) list 8 remnants that show evidence for compact
regions of OH maser emission.
Here, I concentrate on the three remnants W44, IC 443, and 3C391,
which have the most detailed observations.
I use a distance of 3 kpc to W44 (Wolszczan, Cordes, \& Dewey 1991), 1.5 kpc to IC 443 (Fesen \& Kirshner 1980),
and 9 kpc to 3C391 (Reynolds \& Moffett 1993).
The uncertainty in these estimates is $\sim 20-25$ \%, although the distance
to W44 is probably more accurate because of the presence of a pulsar.
In all 3 cases, the remnant radius is $\gsim 6$ pc so that
 the expansion in the interclump
medium of a molecular cloud is expected to have entered the radiative
phase (see \S~3.1).
The general framework for understanding these objects is thus expected to
be a radiative shell interacting with a clumpy molecular gas,
unless the supernova surroundings have been modified by the
progenitor evolution.
The aim of this section is to examine the observational material on 
the remnants to check for
consistency with this framework.

\subsection{W44}

Wootten (1977) identified W44 as a supernova remnant that is associated
with molecular gas.
These  observations did show CO apparently in the vicinity of the remnant,
although they did not conclusively show an interaction (DeNoyer 1983).
However,  Claussen et al.
(1997) have found compact OH masers in the direction of the remnant
(first discovered by Goss \& Robinson 1968) and Seta et al. (1998) have recently found evidence
for  CO clouds that are interacting with the remnant.
The remnant is elongated, with radii of $11\times 15$ pc;  because the
present models are spherically symmetric, I take 13 pc for the radius.

The filled-center, thermal X-ray emission observed from a number of
supernova remnants has led to a model involving the evaporation of clumps
in the postshock region (White \& Long 1991).
The evaporative model of Rho et al. (1994) for the X-ray emission from W44
has an intercloud ambient density of $n_o=0.09-0.26$ cm$^{-3}$,
which is low for a molecular cloud, and the remnant is in an adiabatic phase.
However, the multiwavelength observations of this remnant can be used to
argue for a different type of model.
Koo \& Heiles (1995) briefly discussed a radiative model for this
remnant because they found an $\sim 350\Msun$ HI shell moving at $V_s=150\kms$
that might be identified with the cool shell expected in the radiative
phase.
They also noted that the age of a radiative remnant, $0.3R/V_s=29,000$
years where $R=15$ pc is the remnant radius they chose, is in approximate
agreement with the spindown age of the pulsar, PSR 1853 +01, that
is inside the remnant.
In the model of Rho et al. (1994), the age of the remnant is 
$6,000-7,500$ years, which would require that the pulsar be born
with a period close to its present period of 0.267 s.
However, Koo \& Heiles (1995) did not advocate the radiative model
for W44, because it may not be compatible with the filled-center X-ray emission.

Harrus et al. (1997) preferred the radiative model because the age 
would agree with that of the pulsar if it were born with a period
of 10 msec.
By computing hydrodynamic models, they found that the pulsar age
and the mean remnant radius (13 pc) could be fitted by models
with $E_o=(0.7-0.9)\times 10^{51}$ ergs and $n_o=(3-4)$ cm$^{-3}$.
In these models, the shell velocity was $V_s\sim 120\kms$, somewhat
less than the $150\kms$ found by Koo \& Heiles (1995).
Alternatively, the relationship between $E_o$, $V_s$, $R$, and $n_o$
for radiative supernova remnants found by Chevalier (1974; eq. [46])
yields $n_o=4.7$ cm$^{-3}$ for $E_o=1\times 10^{51}$ ergs,
$V_s= 150\kms$, and $R=13$ pc.
Similar results can be obtained from the computations of
Cioffi et al. (1988).
Harrus et al. (1997) examined the X-ray properties of their radiative
models and found that they could fit the centrally concentrated emission
because of the effect of absorption.
However, the models did show a greater temperature variation than
that indicated by the observations.

Other observations support the view that W44 is in a radiative phase.
Although optical observations are inhibited by dust obscuration,
Giacani et al. (1997) have found H$\alpha$ and [SII] emission that
covers the region of the radio remnant and, in places, shows
a detailed correspondence with the radio structure.
This correspondence is commonly found for radiative shock fronts.
The range of H$\alpha$ surface brightness found by Giacani et al. (1997)
is $1\times 10^{-17}-1\times 10^{-16}$ ergs cm$^{-2}$ s$^{-1}$ arcsec$^{-2}$.
Rho et al. (1994) estimate the reddening to W44 to be $E(B-V)=2.94$,
which leads to an extinction $A_{\lambda}\approx 1.5E(B-V)=4.4$ at the wavelength of
H$\alpha$.
The extinction corrected H$\alpha$ surface brightness is then
 $6\times 10^{-16}-6\times 10^{-15}$ ergs cm$^{-2}$ s$^{-1}$ arcsec$^{-2}$.
In the radiative shock models of Shull \& McKee (1979), the H$\beta$ surface
brightness of a $130\kms$ shock wave moving into a medium with density
$n_o=1$ cm$^{-3}$ is $6.1\times 10^{-6}$ ergs cm$^{-2}$ s$^{-1}$ sr$^{-1}$,
which translates into an H$\alpha$ surface brightness of
$1.7\times 10^{-15}$ ergs cm$^{-2}$ s$^{-1}$ arcsec$^{-2}$ for $n_o=4$ cm$^{-3}$.
The surface brightness can be increased by projection effects when the
shock front is moving close to perpendicular to the line of sight.
The observed H$\alpha$ surface brightness is thus consistent with
the radiative model.

Another observational constraint is the [OI] 63 $\mu$m line emission
from the remnant.   Reach \& Rho (1996) estimate that the observed luminosity in
the line is $\sim 10^3 L_{\odot}=4\times 10^{36}\ergs$.
They suggest that the remnant is in the adiabatic phase because 
 the line loss rate times the age corresponds to the loss of a fraction of $10^{51}$ ergs.
However, the argument requires that the [OI] 63 $\mu$m line carry a
significant part of the total radiative luminosity, which
can be checked by comparison with radiative shock models.
Hollenbach \& McKee (1989) have calculated the emission from radiative
shock fronts for velocities in the range $v_s=30-150\kms$ and $n_o=10^3-10^6$
cm$^{-3}$ and they find that the [OI] 63 $\mu$m surface brightness
is $10^{-6} n_o v_2$ ergs cm$^{-2}$ s$^{-1}$ sr$^{-1}$, where $v_2$
is the shock velocity in units of $100\kms$, for
$n_o v_2 \lsim 10^5$ cm$^{-2}$ s$^{-1}$.
The reason that the line surface brightness is proportional to the
shock particle flux is that the line is responsible for most of the
gas cooling below a temperature of 5,000 K.   % (Hollenbach \& McKee 1989).
The results of Raymond (1979) show that this relation
can be extended down at least to $n_o$ of order unity.
The luminosity in the line is $1.5\times 10^{35} n_o v_2 R_{1}^2\ergs$,
where $R_{1}$ is the radius in units of 10 pc.
This is a small fraction of the total radiative loss.
For $v_2=1.5$, $n_o=4.7$ cm$^{-3}$, and
$R=13$ pc, the expected [OI] 63 $\mu$m luminosity is $2\times 10^{36}\ergs$
The observed luminosity is a factor of 2 higher
than that expected in the radiative model, but this is within the
observational uncertainty considering that only a portion of the
remnant has been observed (Reach \& Rho 1996).
The observed [OI] 63 $\mu$m luminosity thus provides support
for the radiative model.
The peak surface brightness in the [OI] 63 $\mu$m line is
$\sim 10^{-3}$ ergs cm$^{-2}$ s$^{-1}$ sr$^{-1}$ (Reach \& Rho 1996), which 
requires either shock motion into a dense region or projection
effects for an edge-on shock front.
In general, shocks driven into the dense clumps can also be a significant
source of [OI] 63 $\mu$m emission.
As discussed in \S~2, the total mass in clumps in a molecular cloud 
can be several times
the interclump mass, so that most of the mass encompassed by the outer
shock front may be in clumps.
However, the high clump density delays the motion of the shock fronts
into the gas, especially because most of the mass is in the largest clumps.
In addition, coolants other than the [OI] 63 $\mu$m line become important
at densities $\gsim 10^5$ cm$^{-3}$ (Hollenbach \& McKee 1989).
Detailed mapping of the remnant is needed to distinguish the clump
from the interclump shock emission.

The observations of W44 can thus be modeled as a $10^{51}$ erg explosion
in a medium with density $4-5$ cm$^{-3}$ and current shock radius of 13 pc.
The shock front is radiative and the ambient density is at the low
end of the range expected for the interclump density of a molecular cloud.
The interaction with molecular clumps is probably best observed
in the OH maser emission (Claussen et al. 1997), which is likely to
come from a shock front in molecular gas (Elitzur 1976).
An important aspect of these observations is that the Zeeman effect
allows the measurement of the magnetic field strength along the line
of sight, $B_l$.
Claussen et al. (1997) find $B_l\sim 3\times 10^{-4}$ G in W44, which 
is close to the peak field strength expected in the cool shell
($2.5\times 10^{-4}$ G for $n_1=0.47$ and $v_2=1.5$, see \S~3.1).
As discussed in \S~3.3, the pressure in the cooling region of a clump
shock can be higher than that in the radiative shell and the
magnetic field strength is correspondingly increased.

\subsection{IC 443}

IC 443 has strong molecular emission that makes it the clearest
case of supernova interaction with a molecular cloud.
The cloud appears to be in front of the remnant (Cornett, Chin, \& Knapp 1977),
so that the interaction is primarily approaching us.
The molecular cloud found by Cornett et al. (1977) has a mass of only
$\lsim 10^4\Msun$.
This is low for a molecular cloud and, as argued in \S~2, is the type
of cloud that is more likely to show the  interaction of a 
supernova remnant with the molecular material.
Braun \& Strom (1986) noted that the IC 443 complex could be described
by 3 overlapping shells (A, B, and C) with different radii of curvature.
%(5.8, $\sim 10$, and $\sim 19$ pc, respectively).
They attributed this structure to the expansion of the remnant into
shell structures that were created by the actions of stellar winds.
X-ray observations of shell C now show that it is a separate, older supernova
remnant in front of IC 443 (Asaoka \& Aschenbach 1994).
For shells A and B, I take the point of view that they are one remnant;
shell A is expanding into the interclump medium of a molecular
cloud and shell B is the breakout into the low density surroundings.
The breakout morphology is most clearly shown by images at X-ray
(Asaoka \& Aschenbach 1994, Fig. 2) and radio (e.g., Claussen et al. 1997)
wavelengths.
I estimate the radius of shell A to be 7.4 pc.

Shell A comprises about half of the solid angle
subtended by the remnant and includes the bright optical filaments
to the NE of the remnant.
Fesen \& Kirshner (1980) found that the spectra of these filaments
imply an electron density for the [SII] emitting region of 
$\lsim 100$ to 500 cm$^{-3}$ and shock velocities in the range
$65-100\kms$.
If the magnetic field does not limit the  compression to $\sim 10^4$ K,
the preshock  density implied by the higher density filaments is
$10-20$ cm$^{-3}$.
The filaments with a lower density may have their compression limited by
the magnetic field.
The physical parameters are in approximate agreement with those
derived from HI observation.
For an ambient density of $15$ cm$^{-3}$ and a radius of 7.4 pc,
the total swept-up mass is $1000\Msun$ for a half-sphere.
The HI observations of Giovanelli \& Haynes (1979) suggest that there
is $\sim 1000\Msun$ of shocked HI in the shell A part of the remnant,
with velocities up to $\sim 110\kms$.
The application of eq. (46) of Chevalier (1974) for a spherical remnant yields $E=4\times 10^{50}$
ergs for $v_s=100\kms$, $R=7.4$ pc,  and $n_o=15$ cm$^{-3}$.
This energy  probably  underestimates the initial total energy
 because  the breakout into
the low density region decreases the pressure driving the dense shell.

As for W44, the radiative shell should give rise to [OI] 63$\mu$m emission.
This line has been observed from IC 443, but only from a small region
with a molecular clump (Burton et al. 1990).
For $v_s=100\kms$ and $n_o=15$ cm$^{-3}$, the [OI] 63$\mu$m line
luminosity from the half shell should be $\sim 1\times 10^{36}\ergs$.
Mufson et al. (1986) found that the luminosity of the remnant in
the $\it IRAS$ 60$\mu$m band is $\sim 1\times 10^{37}\ergs$.
About 10\% of this band luminosity may be in the [OI]
line.
This line provides an excellent means to investigate the radiative
shell because it is relatively unaffected by dust absorption in the
the cloud that covers the central part of the remnant.

The radiative shell model provides an adequate description of the
observations of shell A.
The model implies an age $t\approx 0.3R/v_s=30,000$ yr for
$r=7.4$ pc and $v_s=100\kms$.
This age is greater than some other estimates based on X-ray observations. 
The brightest region in X-rays is to the N and has a temperature
$\sim 10^7$ K (e.g., Petre et al. 1988), which corresponds to a shock
velocity of $\sim 800\kms$ if it is in an immediate postshock region.
In the current model, the shock reaches the beginning of the
cooling phase at $v_s\approx 600\kms$ when the radius is
4 pc and there is hotter gas in
the interior.
The action of heat conduction may distribute the high interior
heat content so that the temperature is $\sim 10^7$ K despite the
expansion of the radiative shell.
Petre et al. (1988) estimate the X-ray emitting mass to be only
$\sim 10\Msun$.
If this model cannot explain the high temperature, a model involving
interaction with a pre-existing shell may be necessary, so that
the shock front has been moving rapidly until recently.

Recently, Wang et al. (1992) found evidence for gas with a
temperature $\gsim 10^8$ K and concluded that the remnant's age is
$\sim 1,000$ years.
This age has been assumed in other X-ray studies (Asaoka \& Aschenbach 1994;
Keohane et al. 1997).
However, Keohane et al. (1997) present {\it ASCA} observations of the
region and concluded that synchrotron emission provides a more
consistent explanation of the hard X-ray emission, which removes the main
argument for the small age.
Most of the hard X-ray emission comes from a centrally concentrated
source in the southern part of the remnant (Keohane et al. 1997).
This region also has a relatively flat radio spectral index,
$\alpha<0.24$, where flux $\propto \nu^{-\alpha}$ (Green 1986).
These properties point to the presence of a pulsar-driven nebula.
Keohane et al. (1997) argue against this interpretation for a number
of reasons, the strongest of which is that the pulsar would have
to have a velocity $\gsim 5,000\kms$ to reach its position near the
edge of the remnant.
With the age advocated here, the required velocity drops to
$\lsim 200\kms$ and the problem disappears;
IC 443 probably contains a low luminosity pulsar nebula that has
little influence on the shock wave interactions.
Keohane et al. (1997) attribute the emission to energetic electrons
accelerated by the shock front at edge of the remnant and find
another extended hard X-ray region at the eastern edge of the remnant.
A shock with velocity $\sim 100 \kms$, as in the present model,
 is not a promising place to
accelerate electrons to energies $\sim 10$ TeV, but this region is
relatively faint and needs to be observed in more detail.

In a restricted region across the central, eastern part of IC 443,
a rich spectrum of molecular emission has been observed
(van Dishoeck et al. 1993; Richter at al. 1995b, and references therein).
A set of spectroscopically distinct CO clumps (labelled A--G) has been listed by
DeNoyer (1979b) and Huang, Dickman, \& Snell (1986).
These clumps, with  sizes $\sim$1 pc, have masses of $3.9-41.6\Msun$ as
deduced from  $^{12}$CO lines (Dickman et al. 1992).
The corresponding densities, $n_H\lsim 500$ cm$^{-3}$, are low but
should be regarded as lower limits because the $^{12}$CO emission is
assumed to be optically thin.
Analysis of absorption in one place yielded a preshock density of
$n({\rm H_2})\approx 3,000$ cm$^{-3}$ (van Dishoeck et al. 1993).
The properties of these clumps are consistent with those seen in
quiescent molecular clouds (see \S~2).
%Velocity variations across clumps have been interpreted in terms of
%ablation (Dickman et al. 1992; Tauber et al. 1994).
Especially high spatial resolution is possible in the H$_2$ infrared
lines (Richter et al. 1995b), which show clumps down to a scale of
1$^{\prime\prime}$ ($2\times 10^{16}$ cm$=0.007$ pc at a distance of 1.5 kpc) but not smaller.
Richter et al. (1995b) note that the emission has a knotty appearance
that is unlike the filamentary appearance of the optical emission in IC 443
and the H$_2$ emission in the Cygnus Loop.
This type of structure is consistent with the fractal structure of
molecular clumps deduced by Falgarone \& Phillips (1991).

The interpretation of the molecular line strengths is still
controversial, but the emission is most consistent with that from a partially
dissociating $J$(jump)-type shock (Burton et al. 1990;
van Dishoeck et al. 1993; Richter at al. 1995a,b).
The required shock velocity, $25-30\kms$, is consistent with the
velocities observed in the strongest emission.
The dissociated H$_2$ can be observed as a high column density of
shocked HI at the positions of the clumps (Braun \& Strom 1986).
For $v_c=25-30\kms$ and $v_{rs}=100\kms$, $\rho_{cl}/\rho_{rs}$ must
be in the range $5.4-9.0$ according to eq. (21).
The density ratio can vary significantly with relatively little
effect on the shock velocity.
The clump shock velocity is consistent with $n_{rs}=500$ cm$^{-3}$
and $n_{cl}=3000$ cm$^{-3}$, which are plausible values.
The ratio of postshock pressure in the clump to that in the
interclump medium (with $n_o=15$ cm$^{-3}$) is 18, which is different
from the ram pressure equilibrium typically found in cloud interactions
in adiabatic supernova remnants (McKee \& Cowie 1975).
Moorhouse et al. (1991) found a high pressure in IC 443 from H$_2$
observations and proposed an explanation similar to that given here.
I have shown how the required radiative shell follows naturally from
the global properties of the supernova remnant.

The column density through the radiative shell is
$N_H\approx n_o R/3\approx 10^{20}$ cm$^{-2}$.
For $n_{cl}=3000$ cm$^{-3}$,
the column density through the clumps is $10^{22} \ell_{\rm pc}$ cm$^{-2}$,
where $\ell_{\rm pc}$ is the path length through the clump in pc.
The larger observed clumps are thus expected to be passed over by the
radiative shell and left in the interior of the remnant (see \S~3.3).
For clumps with a size $\lsim 0.02$ pc, the shock front breaks out of
the clump first and there is the possibility of further acceleration
by the radiative shell.
The initial shock front through the clump may not dissociate
molecules, which are then accelerated by the shell.
Low column densities of high velocity molecular gas can be produced
in this way.
High velocity molecular gas has been observed in IC 443 at the edge
of one of the clumps (Tauber et al. 1994).

\subsection{3C391}

3C391 is another remnant that is likely to be interacting with
a molecular cloud, based on the apparent breakout morphology at
radio and X-ray wavelengths (Reynolds \& Moffett 1993; Rho \& Petre 1996)
and on the detection of OH maser emission (Frail et al. 1996).
At a distance of 9 kpc, the bright radio emission to the NW, which
is presumably the cloud interaction, has a radius of curvature
of 6 pc.
This radius is just bigger than that of the transition to the
radiative phase for an interclump density of 10 cm$^{-3}$ (see \S~3.1).
The dynamical status of the remnant thus depends on the exact
interclump density, but if the remnant has made a transition to the
radiative phase, the shock velocity may still be 100's of km s$^{-1}$.
The remnant has not been detected at optical wavelengths; Reynolds
\& Moffett (1993) attribute this to the large absorption along the
line of sight.
The [OI] 63$\mu$m line is not much affected by absorption.
It has been detected in two places in 3C391 and the whole remnant is inferred
to have a line luminosity $\sim 10^3~L_{\odot} = 4\times 10^{36}\ergs$
(Reach \& Rho 1996).
In the shock model for the line, the luminosity for a spherical remnant is
$L_{\rm [OI]}=1.5\times 10^{36} n_1 v_2 R_1^2\ergs$, where $R_1$ is
the remnant radius in units of 10 pc.
The line luminosity for $R_1=0.6$ and $v_2=3$ is thus
$L_{\rm [OI]}=1.6\times 10^{36} n_1\ergs$, which is in adequate
agreement with the observations, considering that only part of the
remnant was observed. % and that the radiative shock wave probably covers
%only part of the remnant.
Although the spectral resolution available to Reach \& Rho (1996)
was poor, they found that the velocity centroid of the [OI] line
was shifted by $\sim 400\kms$ between the two places that they detected
the line.
This velocity difference is suggestive of the velocities in newly
formed radiative shock fronts, which are faster than the shocks
in molecular clumps.

Reach \& Rho (1998) have recently found that there is strong 
molecular emission
from a shocked  clump in the southern part of the remnant.
They note that the CO line brightness implies a pressure $\sim 10^2$ higher
than that expected for a $10^{51}$ erg supernova.
As discussed in \S~3.3, an overpressure of this order can be generated
if a radiative shell interacts with a dense clump.

\subsection{X-ray Emission and Heat Conduction}

The X-ray observations of the 3 remnants discussed in \S~4 present
the greatest challenge to the type of model proposed here.
In the standard model for a radiative supernova remnant, there
is hot gas left at the center.
However, the hot gas has its highest density close to the shell
and the temperature increases toward smaller radii (Chevalier 1974;
Cioffi et al. 1988; Harrus et al. 1997).
The result is the expectation of a shell X-ray source with a decreasing
temperature toward larger radii.
The observations of these remnants show centrally concentrated
thermal X-ray emission with little temperature variation (Rho et al. 1994
on W44; Petre et al. 1988 on IC 443; Rho \& Petre 1996 on 3C391).
Harrus et al. (1997) have attempted to solve this problem for
W44 by suggesting that the shell emission is absorbed, leaving the
centrally concentrated emission.

These remnants belong to a larger class of remnants that are
characterized by their %X-ray emission.
%During the past decade, observers have discovered a growing class of 
%supernova remnants with 
center-filled, thermal X-ray emission.
Their X-ray characteristics have been used to advocate 
 the evaporating cloud model of White \& Long (1991).
In this model, which is an elaboration of the original evaporative
model of McKee \& Ostriker (1977), 
the main supernova shock front propagates without losing energy
through an intercloud medium.
Embedded in this medium are dense clouds that are not accelerated by
the passage of the blast wave.
The clouds find themselves in a hot medium, which leads to evaporation
of the cloud material by saturated heat conduction.
White \& Long (1991) found that if the mass evaporation rate has a $t^{-1}$
time dependence, the flow can be described by a self-similar solution.
If the evaporation time is short, the clouds are completely evaporated
close to the shock front.
If the evaporation time is very long, the blast wave is not affected by
the clouds.
But there is an intermediate regime in which the cloud evaporation adds
mass to the center of the supernova remnant and can give a center-filled
remnant.
The remnants are taken to be in this regime and most of the X-ray
emission comes from evaporated clump gas.
Applications of this model lead to  low values for the
preshock intercloud H density: 0.09--0.26 cm$^{-3}$ in W44 (Rho et al. 1994),
0.039 cm$^{-3}$ in W28 (Long et al. 1991) which is another source
of OH maser emission (Claussen et al. 1997), 
and 0.07--0.23 cm$^{-3}$ in 3C391 (Rho \& Petre 1996).

In both the present model and the evaporative model, the remnants
are interacting with a clumpy medium, but the nature of the interclump
medium is very different in the two models.
The discussion in this section provides strong arguments that the
3 remnants under consideration are interacting with a dense ambient
medium and have developed radiative shells.
There are 2 reasons why clump evaporation is unlikely to be an
important process in the present model.
First, evaporation leads to a higher interior density, which helps
with modeling the X-ray structure.
However, the cooling time is shortest in the interior, so that the
radiative shell forms well inside of the outer shock front, as
can be seen in the numerical calculations of Cowie, McKee, \& Ostriker (1981).
This is incompatible with the observations.
The other reason is that the estimated mass of X-ray emitting gas 
is much less than the estimated mass of the radiative shells.
Clump evaporation may be taking place, but not at a level that is
important for the evolution.

Although clump evaporation is probably not significant, heat conduction
in the hot interior gas probably is significant.
Harrus et al. (1997) found that their non-conduction models were
unable to fit the isothermal nature of the X-ray emission from W44.
Conduction in the interclump gas is plausible because stellar polarization
observations of molecular clouds show that this gas has a significant
uniform component to the magnetic field (see \S~2).
Models for radiative remnants with heat conduction are presented by
Cui \& Cox (1992) and Shelton, Smith, \& Cox (1995) noted that
heat conduction without evaporation could be responsible for the
filled center appearance of W44.
The conditions here are such that conduction does not impede formation
of the radiative shell, but can operate in the hot interior.

\section{NONTHERMAL PARTICLES AND THEIR EMISSION}

The presence of relativistic electrons in supernova remnants like those
discussed in \S~4 has long been inferred from their radio emission.
A recent development has been the likely detection of a number of
supernova remnants in energetic $\gamma$-rays, including W44, IC 443,
and W28 (Esposito et al. 1996).
Possible sources of the emission are  bremsstrahlung and inverse Compton
radiation associated with energetic electrons and pion decays
associated with energetic protons.
These emission mechanisms have been extensively studied in the
context of the supernova remnant emission (de Jager \& Mastichiadis 1997;
Sturner et al. 1997;
Gaisser, Protheroe, \& Stanev 1998).
Although these studies have treated the emission mechanisms in
detail, their treatments of the supernova remnant structure 
are more rudimentary.
For example, Sturner et al. (1997) present a model for IC 443 in
which the interior density is constant at 4 times the ambient density
of 10 cm$^{-3}$ and the magnetic field is constant at 4 times the
ambient value of 5 $\mu$G.
They assume that particle acceleration terminates when a remnant enters
the radiative phase, but have IC 443 as non-radiative at an age of
5,000 yrs.
Here, I examine the nonthermal component in the context of the 
supernova remnant models
developed in the previous section.

The nonthermal component is observed at the highest spatial resolution 
at radio wavelengths.
A close correlation between the
radio and optical structure has been found in IC 443 (Duin \& van der Laan 1975)
and in W44 (Giacani et al. 1997).
The optical structure in IC 443 and other radiative shock front regions
is best explained as emission from a wavy sheet.
The fact that the radio emission shows a similar structure implies that
a major part of it is associated with a thin region near the radiative shock.
Duin \& van der Laan (1975) attribute the emission to the compressed 
ambient magnetic field and cosmic ray electrons.
Particle acceleration at the shock front can also be a factor
(e.g., Blandford \& Cowie 1982).

Radiative shock fronts are also driven into the dense clump gas
and the high pressure should result in a high magnetic field strength
in the postshock region (\S~3.3).
However, the clumps do not appear to be significantly enhanced sources of
radio synchrotron emission.
Figures 2 and 3 of Claussen et al. (1997) show that except for 
the sources OH D and OH E in W44, there is little correlation between
the compact OH masers and the radio continuum emission in W44 and IC 443.
In IC 443, the small region of OH masers does match up with a CO
clump (Claussen et al. 1997), but the radio continuum emission shows
little correlation with the CO clumps (Dickman et al. 1992).
The absence of strong emission is not surprising because the shock fronts
into the clumps are not ionizing shock fronts.
Under these circumstances, Alfven waves are damped in the neutral
gas and energetic particles may be able to escape from the shock region.
%the standard first-order Fermi acceleration cannot take place.

The conclusion is that the radio emission is from the compressed
regions behind the radiative shock fronts in the interclump gas.
A possible problem with this source is free-free absorption by
compressed gas in the cooling, recombining region.
Erickson \& Mahoney (1985) have pointed out the lack of free-free
absorption in IC 443 down to $\sim 20$ MHz, which may be
a problem in a radiative shock model.
At $10$ MHz, absorption does appear to be present.
I estimate the free-free optical depth from the intensity of
$H\beta$ emission expected from a radiative shock front.
Taking the $H\beta$ intensity from Raymond's (1979) shock models
for a $100\kms$ shock and
the $H\beta$ emissivity coefficient from Osterbrock (1989),
I find the free-free depth at $T\approx 8,000$ K to be
\begin{equation}
\tau_{ff}\approx 0.01 \nu_8^{-2} n_1,
\end{equation}
where $\nu_8$ is the frequency in units of $10^8$ Hz.
For IC 443, the cooling region is just becoming optically thick
at $10-20$ MHz, which is roughly compatible with the observations.
The H$\beta$ intensity observed by Fesen \& Kirshner (1980) is
higher than that in the shock model by $\gsim$ 10 in some bright
filaments, which can be
attributed to the fact that bright filaments are probably shock
fronts viewed edge-on.
%(Check Ha surface brightness of 443)

The observed radio flux from IC 443, $\sim 160$ Jy at 1 GHz (Erickson \& Mahoney 1985), can be used to find the
energy density of the relativistic electrons in the context of
the radiative shell model.
The emitting volume of the thin shell is $4\pi R^2\Delta 
R\approx(4\pi R^3/3)(\rho_o/\rho_{sh})$, where $(\rho_o/\rho_{sh})=15/500$.
With $R=7.4$ pc, the emitting volume is $1.5\times 10^{57}$ cm$^3$.
I take $B\sin\theta =2\times 10^{-4}$ G in the thin shell, so that the
emitting electrons at 1 GHz have energies $E\approx 0.56$ GeV.
The radio spectral  index is observed to be $\alpha=0.36$ (Erickson
\& Mahoney 1985), where $\alpha$ is defined by flux $\propto \nu^{-\alpha}$;
this corresponds to an energy spectral index $\Gamma=1.72$ 
($N(E)dE\propto E^{-\Gamma} dE$).
The basic parameters given here for the radiative shell of IC 443 are close to those
given by Duin \& van der Laan (1975), and the radio flux can  again be
attributed to the compression and acceleration
of ambient cosmic ray electrons in the
dense shell.
Blandford \& Cowie (1982) discussed particle acceleration in older
remnants with radiative shock waves, like IC 443.
They noted that very efficient cosmic ray acceleration at the shock
front cannot be occurring, or the high compression in the radiative
region of the shock front could not take place.
However, in their model the ambient cosmic ray electrons are
accelerated in the shock front by the first-order Fermi mechanism,
which produces a $\Gamma=2$ energy spectrum.
This is significantly steeper than the spectral index observed in IC 443.
If the emitting electrons are compressed ambient cosmic rays,
the spectral index may depend on that of the ambient cosmic rays.
As a first approximation, I take the cosmic ray spectrum in the cloud to
be the same as that in the Galaxy.
Observations of molecular clouds with {\it CGRO} indicate that the
cosmic ray content of molecular clouds is similar to that in the
rest of the interstellar medium (Digel et al. 1995, 1996).
Taking into account the compression in the radiative shock front,
the initial preshock electron energy range covered by the 30 MHz -- 10 GHz
observations of IC 443 is 50 -- 900 MeV.
Longair (1994, p. 284) estimates
$\Gamma=1.6$ over the energy range 10--100 MeV and $\Gamma=1.8$ over 100 Mev -- 1 GeV for ambient, interstellar cosmic ray electrons.

For shock acceleration in the test particle limit (Blandford \& Ostriker
1978), an input spectrum with $\Gamma>2$ results in $\Gamma=2$
in the postshock region.
If $\Gamma<2$ initially, the particle spectrum is preserved in the
postshock region.
The relatively flat ambient cosmic ray spectrum is thus preserved
in the postshock region and can explain the flat radio synchrotron
spectrum that is observed in IC 443.
This spectrum would be difficult to understand if electrons were
injected at the shock front.
Longair (1994) attributes the flat spectrum of the ambient cosmic rays
to Coulomb losses in the interstellar medium, which he estimates
as $10 n$ eV yr$^{-1}$ where $n$ in the gas density in cm$^{-3}$.
Coulomb losses in the dense shell of the supernova remnant are also
possible.
For particle energy $E=500$ MeV and $n=500$ cm$^{-3}$, the loss time is
$E(dE/dt)^{-1}=10^5$ yr.
This is longer than the estimated age of the remnant ($3\times 10^4$ yr),
but is suffiently close that Coulomb losses in the shell could be
a factor for some remnants.

The electron particle spectrum required to produce the observed flux from
IC 443, given the above spectrum and volume, is 
$N(E)=N_o E^{-\Gamma}=3\times 10^{-11}E^{-1.72}$ in cgs units.
On the assumption that the particle pitch angles are isotropized
during the postshock compression, the individual particles gain energy as
$s^{1/3}$, where $s$ is the compression factor, and 
$N_o\propto s^{(2+\Gamma)/3}$.
From the theory of Blandford \& Ostriker (1978), the shock acceleration
of the ambient cosmic rays leads to an increase in $N_o$ by a factor
of 14.
The combination of these factors implies that the electron spectrum
ahead of the shock front is $N(E)=2\times 10^{-14} E^{-1.72}$.
This spectrum compares adequately with the spectrum $N(E)=2\times 10^{-14} E^{-1.6}$
given by Longair (1994) for ambient cosmic ray electrons up to $\sim 100$ MeV;
at 100 MeV, the required spectral density is 3 times that provided by
the ambient cosmic rays.
In the present model, the electron energy density in the dense shell
is $\sim 1$ \% of the magnetic energy density.
I have neglected the structure of the radiative shell, 
 but this should not significantly affect the results.
The present model for the radio emission
resembles that of Blandford \& Cowie (1982) except for the difference in
spectral index and in the overall picture.
Blandford \& Cowie take the emission to be from crushed interstellar clouds,
whereas here it is from the radiative shell in the interclump medium of a molecular cloud.

Further support for the molecular cloud model comes from polarization
observations of the radio synchrotron emission.
Kundu \& Velusamy (1972) found that both IC 443 and W44 show an
ordered field direction and that the polarization in W44 is up
to 20\%, which is unusually high for a supernova remnant.
In the case of IC 443, the magnetic field direction deduced from the
synchrotron emission lines up with that deduced from interstellar
polarization in the surrounding cloud.
The emission is thus consistent with the compressed magnetic field coming
from the interclump component of a molecular cloud with a significant
uniform magnetic field (see \S~2).

The relativistic electrons that give rise to synchrotron radiation
can also give rise to bremsstrahlung radiation.
The bremsstrahlung intensity for a power law electron spectrum is
$1.0\times 10^{-15} n N_o E^{-\Gamma}/(\Gamma-1)$ cm$^{-3}$ ergs$^{-1}$
s$^{-1}$ (Longair 1994, p. 269).
If electrons of an energy $E$ can be considered to give synchrotron
radiation at a frequency $\nu_s$ and bremsstrahlung radiation at
energy $E$, then the ratio of bremsstrahlung power to synchrotron
power for those electrons is
\begin{equation}
{(\nu L_{\nu})_{\rm brems}\over (\nu L_{\nu})_{\rm syn}}=
10\left(n\over 500 {\rm ~cm^{-3}}\right)
\left(B\sin\theta\over 2\times 10^{-4} {\rm~G}\right)^{-3/2}
\left(\nu_s\over 1 {\rm~GHz}\right)^{-1/2},
\end{equation}
where the bremsstrahlung emission is at photon energy $E$,
the radio synchrotron emission is at frequency $\nu_s$,
and $\Gamma=2$ has been assumed.
A similar calculation for $\Gamma=1.72$ gives $\sim 11$ for the ratio
when the reference values are used.
The critical frequency for synchrotron emission from an electron of energy $E$
is
\begin{equation}
\nu_{\rm syn}=3.2\left(B\sin\theta\over 2\times 10^{-4} {\rm~G}\right)
\left(E\over {\rm GeV}\right)^2 \qquad {\rm GHz},
\end{equation}
so for IC 443, an electron radiating at 1 GHz has $E=0.56$ GeV.
The reference values for IC 443, used in eq. (33), then give the
ratio of $\gamma$-ray power (at 0.56 GeV) to radio synchrotron power
(at 1 GHz) for the remnant.
The observed value for the ratio is about 30 (Esposito et al. 1996; 
see fig. 11 of Sturner et al. 1997).
The current model thus gives the correct ratio to order of magnitude,
although the model ratio does fall short of the observed ratio.
One possibility is that there is a contribution to the
bremsstrahlung emission from the shocked
molecular clump material; the high density in this gas ($n_H\gsim 10^5$
cm$^{-3}$) favors bremsstrahlung emission over synchrotron emission.
Sturner et al. (1997) also propose a bremsstrahlung model for the high
energy $\gamma$-ray emission from IC 443, but in the context of
a different supernova remnant model (uniform $n=40$ cm$^{-3}$ and
$B=2\times 10^{-5}$ G inside the remnant).

Other possibilities for the high energy $\gamma$-ray emission are
inverse Compton emission and pion decays triggered by the collisions
of relativistic protons with thermal gas.
The importance of inverse Compton compared to bremsstrahlung
 depends on the ratio of the radiation energy density to
the gas density.
Gaisser et al. (1998) have discussed the radiation energy density
in IC 443. % and find that it is $\lsim 1$ eV cm$^{-3}$.
For $n=500$ cm$^{-3}$ taken here, the inverse Compton
mechanism is not a factor.
The spectrum expected for pion decay $\gamma$-ray emission does not
fit the observed $\gamma$-ray spectrum (Esposito et al. 1996; 
Sturner et al. 1997; Gaisser et al. 1998) so that bremsstrahlung
emission must be a significant contributor to the emission in the
present model.
Pion decay emission may still make a contribution to the $\gamma$-rays,
but a detailed calculation of the emission is beyond the scope of this
paper.
The present model does give a definite scenario (acceleration and
compression of ambient cosmic rays) that can be examined.

The nonthermal emission from the other supernova remnants can
be investigated from the same point of view.
The radio spectral index of W44 is $\alpha=0.33$ (Kovalenko,
Pynzar, \& Udal'tsov 1994), which can
again be explained in the present model by the acceleration and
compression of ambient cosmic ray electrons.
Because of the flat spectral index, de Jager \& Mastichiadis (1997) suggest
a model in which the electrons come from the pulsar associated with
the remnant.
The ability of the electrons to move into a shell with a tangential
magnetic field is not clear.
In the case of 3C391, the spectral index is steeper, $\alpha\approx
0.55$ (Goss et al. 1979).
One possibility is that this remnant has recently made the transition to
being radiative in the interclump medium and there is an important 
contribution to the relativistic electrons from particles that have
been injected into the shock acceleration process during the earlier phase.

The main point here is not to present definitive models for the
remnants, but to give a supernova remnant model that can be used
in more detailed treatments of the various types of emission.
The approximate results appear promising for more detailed
calculations.

\section{DISCUSSION}

The aim of this paper has been to show that observations of a number
of luminous supernova remnants can be interpreted in terms
of interaction with a clumpy molecular cloud.
The remnants become radiative in the interclump medium and the
radiative shells then interact with molecular clumps.
The magnetically supported radiative shells can be observed as cool HI shells and
by the radiative shock emission.
Because the extinction is frequently high to these remnants in
dense regions, the study of infrared line emission can be especially
fruitful.
The [OI] 63$\mu$m line has already been found to be a powerful
probe of these remnants.
A prediction of this model is that the high-energy $\gamma$-ray
emission detected from some of these remnants is from the
radiative shell.
This prediction can eventually be checked with experiments that have
a higher spatial resolution than is currently possible, such as
{\it GLAST} ({\it Gamma-ray Large Area Space Telescope}).
A more detailed treatment of the expected emission spectrum for 
acceleration of ambient cosmic ray should also be carried out.

The interaction of the radiative shell with molecular clumps
can give a number of features observed from these remnants, such as
higher pressures than those in the radiative shell and the
acceleration of molecules to relatively high velocities.
Observations of molecular clouds have shown a great deal of small
scale structure in the molecular material and this can be seen
directly in H$_2$ observations of the remnant IC 443 (Richter et al. 1995b).
There may also be instabilities associated with the interaction.
Numerical computations of the interaction of a radiative shell with
a dense clump should be carried out to elucidate the hydrodynamic
features of the interaction.
These calculations will be helpful in the interpretation of detailed
molecular shock wave observations.

The hot interiors of the remnants are a source of X-ray emission.
The observations show that the emission is close to isothermal,
which suggests the action of heat conduction.
However the evaporation of clumps does not appear to be a dominant
process because of the small X-ray emitting mass and the fact that
the radiative shells appear to be at the outer edges of the remnants.
Evolutionary calculations that include heat conduction are needed
for a detailed comparison with these remnants.
After this paper was submitted, the paper by Shelton et al. (1998)
on a detailed model for W44 became available.
Shelton et al. (1998) present a model that is similar to the one
presented here in that W44 is found to be a radiative remnant in a
medium with density $\sim 6$ cm$^{-3}$, the synchrotron emission is
from compressed ambient cosmic ray electrons (although they do not
include shock acceleration of these electrons), and the [OI] and H$\alpha$
emission are from the radiative shock front.
They go beyond the present work in showing how expansion into a medium
with a density gradient can provide a detailed fit to the HI observations
(Koo \& Heiles 1995) and in calculating the X-ray emission for a hydrodynamic
model with heat conduction.
The model is significantly more successful in fitting the observed
X-ray emission than the non-conduction
model of Harrus et al. (1997), although the X-ray emission is still not
as centrally peaked as is observed.
A difference with the present model is that the OH maser emission is
attributed to emission from the radiative shock front and Shelton et
al. (1998) find no compelling evidence for interaction with a molecular
cloud.
The point of view taken here that the OH masers are associated with shock
fronts in molecular clumps is supported by the recent observations of
Frail \& Mitchell (1998) who find that there are CO emitting clumps
that are associated with the OH maser emission in W44.

Massive stars are expected to frequently have sufficient photoionizing
radiation and winds to sweep out a region $\gsim 15$ pc in size
around the stars.
It is only the massive stars at the low mass end ($\lsim 12\Msun$) that
may interact directly with a molecular cloud.
Although these are a minority of the remnants, they are bright because
of the interaction with dense gas and so stand out in the 
observational sample.
The further investigation of these objects can be expected to shed light
on a number of physical processes, from molecular shock emission to
the high-energy $\gamma$-ray emission from cosmic rays.

\acknowledgments

I am grateful to David Bertsch, Claes Fransson, and Robin Shelton for correspondence
and discussion, and to the referee, Christopher McKee, for perceptive
comments on the manuscript.
The supernova remnant web site of David Green 
(http://www.mrao.cam.ac.uk/surveys/snrs/)
was a useful guide to the observations.
Support for this work was provided in part by NASA grant NAG-5-3057.

\clearpage

\begin{figure}
%\plotone{sgi9259.eps}
\caption{The structure of the radiative shell for the case $B_t\propto\rho r$.
The shock front is at $\zeta=0$.
The density and pressure are proportional to $E_o$ and $P_o$, respectively,
and $W_o$ is a measure of the deviation of the velocity from a constant.
The expansion exponent of the radiative shell is $\eta=0.3$.}
\caption{The same as fig. 1, except for the case $B_t\propto\rho$.}
\caption{Schematic figure of the interaction of a radiative shell,
moving at velocity $v_{rs}$,
with a molecular clump.
The interaction generates a dense slab bounded by shock waves
(dashed lines) and
moving at velocity $v$.
The densities of the molecular clump, $\rho_{cl}$, the radiative
shell, $\rho_{rs}$, and the interclump medium, $\rho_{o}$,
are indicated.}
\caption{The evolution of the dimensionless position $\zeta_s$, velocity
$U_o$, and column density $\Omega$ as a function of dimensionless
time $y$ for a slab of gas driven into a molecular clump by a 
radiative shell.
The dashed line is an analytical approximation to $\Omega$ that becomes
increasingly accurate at late times.
By the point $\zeta_s=0.453$, the reflected shock front has moved through
the radiative shell and the slab continues to move by its momentum.}
\end{figure}

\clearpage
\begin{table*}

\noindent{Table 1. Radii of HII Regions and Wind Bubbles}

\vspace{1cm}

\begin{tabular}{clrccc}
\hline
Mass & Type &  $R_{\rm ionized}$ & $\dot M$  & $\tau_{ms}$ & $R_b$ \\
($\Msun$) &   &  (pc)  &  $\mll$  & (yr)  &   (pc) \\
\hline
20  &  O9V  & 13.6 & $1\times 10^{-7}$  & $7\times 10^6$  & 11  \\
16  &  B0V  & 8.0 &  $6\times 10^{-8}$ &  $9\times 10^6$ &   10 \\
12  &  B1V  & 1.6 &  $6\times 10^{-9}$ &  $13\times 10^6$ &  5.3  \\
10  &  B2V  & 1.0 & $5\times 10^{-10}$  & $18\times 10^6$  &  2.6  \\
8  &  B3V  & 0.5 &  $1\times 10^{-11}$ &  $26\times 10^6$ &  0.8  \\
 \hline
\end{tabular}
\end{table*}


\clearpage



\begin{thebibliography} {}

\bibitem[]{}
Asaoka, I., \& Aschenbach, B. 1994, \aap, 284, 573

\bibitem[]{}
Bertoldi, F., \& McKee, C. F. 1990, \apj, 354, 529

\bibitem[]{}
Bertschinger, E. 1986, \apj, 304, 154

\bibitem[]{}
Blandford, R. D., \& Cowie, L. L. 1982, \apj, 260, 625


\bibitem[]{}
Blandford, R. D., \& Ostriker, J. P. 1978, \apj, 221, L29

\bibitem[]{}
Blitz, L. 1993, in Protostars and Planets III, ed. E. H. Levy and J. I. Lunine (Tucson: Univ. of Arizona), 125

\bibitem[]{}
Braun, R., \& Strom, R. G. 1986, \aap, 164, 193

\bibitem[]{}
Burton, M. G., Geballe, T. R., Brand, P. W. J. L., \& Webster, A. S. 1988, MNRAS, 231, 617

\bibitem[]{}
Burton, M. G., Hollenbach, D. J., Haas, M. R., \& Erickson, E. F. 1990, ApJ, 355, 197

\bibitem[]{}
Chevalier, R. A. 1974, ApJ, 188, 501

\bibitem[]{}
Cioffi, D. F., McKee, C. F., \& Bertschinger, E. 1988, ApJ, 334, 252

\bibitem[]{}
Claussen, M. J., Frail, D. A., Goss, W. M., \& Gaume, R. A. 1997, 
ApJ, 489, 143

\bibitem[]{}
Cornett, R. H., Chin, G., \& Knapp, G. R. 1977, \aap, 54, 889

%\bibitem[]{}
%Cowie, L. L., \& McKee, C. F. 1977, ApJ, 211, 135

\bibitem[]{}
Cowie, L. L.,  McKee, C. F., \& Ostriker, J. P. 1981, ApJ, 247, 908

\bibitem[]{}
Cox, D. P. 1972, \apj, 178, 159

\bibitem[]{}
Cui, W., \& Cox, D. P. 1992, \apj, 401, 206

\bibitem[]{}
de Jager, O. C., \& Mastichiadis, A. 1997, ApJ, 482, 874

\bibitem[]{}
de Jager, C., Nieuwenhuijzen, H., \& van der Hucht, K. A. 1988, A\&AS, 72, 259

\bibitem[]{}
DeNoyer, L. 1979a, ApJ, 228, L41

\bibitem[]{}
DeNoyer, L. 1979b, ApJ, 232, L165

\bibitem[]{}
DeNoyer, L. 1983, ApJ, 264, 141

\bibitem[]{}
Dickman, R. L., Snell, R. L., Ziurys, L. M., \& Huang, Y.-L. 1992, ApJ, 400, 203

\bibitem[]{}
Digel, S. W., Hunter, S. D., \& Mukherjee, R. 
1995, \apj, 441, 270

\bibitem[]{}
Digel, S. W., Grenier, I. A., Heithausen, A., Hunter, S. D., \& Thadeus, P.
1996, \apj, 463, 609

\bibitem[]{}
Draine, B. T., \& McKee, C. F. 1993, ARA\&A, 31, 373

\bibitem[]{}
Duin, R. M., \& van der Laan, H. 1975, \aap, 40, 111

\bibitem[]{}
Elitzur, M. 1976, ApJ, 203, 124

\bibitem[]{}
Erickson, W. C., \& Mahoney, M. J. 1985, \apj, 290, 596

\bibitem[]{}
Esposito, J. A., Hunter, S. D., Kanbach, G., \& Sreekumar, P. 1996, ApJ, 461, 820

\bibitem[]{}
Falgarone, E., \& Phillips, T. G. 1991, in Fragmentation of Molecular Clouds and Star Formation, ed. E. Falgarone, F. Boulanger, \& G. Duvert (Dordrecht: Kluwer), 119

\bibitem[]{}
Fesen, R. A., \& Kirshner, R. P. 1980, \apj, 242, 1023

\bibitem[]{}
Frail, D. A., Goss, W. M., Reynoso, E. M., Giacani, E. B., Green, 
A. J., \& Otrupcek, R. 1996, AJ, 111, 1651

\bibitem[]{}
Frail, D. A., \& Mitchell, G. F. 1998, \apj, submitted (astro-ph/9807011)

\bibitem[]{}
Gaisser, T. K., Protheroe, R. J., \& Stanev, T. 1998, ApJ, 492, 219

\bibitem[]{}
Giacani, E. B., et al. 1997, AJ, 113, 1379

\bibitem[]{}
Giovanelli, R., \& Haynes, M. P. 1979, ApJ, 230, 404

\bibitem[]{}
Goss, W. M., \& Robinson, B. J. 1968, Ap. Lett., 2, 81

\bibitem[]{}
Goss, W. M., Skellern, D. J., Watkinson, A., \& Shaver, P. A. 1979, A\&A,
78, 75

\bibitem[]{}
Green, D. A. 1986, \mnras, 221, 473

%\journa{Hester, J. J., et al.}{1996}{AJ}{111}{2349}

\bibitem[]{}
Harrus, I. M., Hughes, J. P., Singh, K. P., Koyama, K., \& Asaoka, I. 1997,
ApJ, 488, 781

\bibitem[]{}
Heiles, C., Goodman, A. A. McKee, C. F., \& Zweibel, E. G. 1993, in Protostars and Planets III, ed. E. H. Levy and J. I. Lunine (Tucson: Univ. of Arizona), 125

\bibitem[]{}
Hollenbach, D., \& McKee, C. F. 1989, ApJ, 342, 306

\bibitem[]{}
Huang, Y.-L., Dickman, R. L., \& Snell, R. L. 1986, ApJ, 302, L63

\bibitem[]{}
Keohane, J. W., Petre, R., Gotthelf, E. V., Ozaki, M., \& Koyama, K.
1997, \apj, 484, 350

\bibitem[]{}
Koo, B.-C., \& Heiles, C. 1995, ApJ, 442, 679

\bibitem[]{}
Kovalenko, A. V., Pynzar, A. V., \& Udal'tsov, V. A. 1994, Astron. Rep., 38, 95

\bibitem[]{}
Kramer, C., Stutzki, J., R\"ohrig, R., \& Corneliussen, U. 1998, \aap,
329, 249

\bibitem[]{}
Kundu, M. R., \& Velusamy, T. 1972, \aap, 20, 237

\bibitem[]{}
Long, K. S., Blair, W. P., White, R. L., \& Matsui, Y. 1991, ApJ, 373, 567

\bibitem[]{}
Longair, M. S. 1994, High Energy Astrophysics, 2nd Ed., Vol. 2 (Cambridge:
Cambridge Univ. Press)

\bibitem[]{}
McKee, C. F., \& Cowie, L. L. 1975, ApJ, 195, 715

\bibitem[]{}
McKee, C. F., Hollenbach, D. J., Seab, C. G., \& Tielens, A. G. G. M.
 1987, ApJ, 318, 674

\bibitem[]{}
McKee, C. F., \& Ostriker, J. P. 1977, ApJ, 218, 148

\bibitem[]{}
McKee, C. F., Van Buren, D., \& Lazareff, B. 1984, ApJ, 278, L115

\bibitem[]{}
Moorhouse, A., Brand, P. W. J. L., Geballe, T. R., \& Burton, M. G. 1991,
\mnras, 253, 662

\bibitem[]{}
Mufson, S. L., McCollough, M. L., Dickel, J. R., Petre, R., White, R., \& Chevalier, R. 1986, AJ, 92, 1349

\bibitem[]{}
Myers, P. C., \& Goodman, A. A. 1991, \apj, 373, 509

\bibitem[]{}
Osterbrock, D. E. 1989, Astrophysics of Gaseous Nebulae and Active
Galactic Nuclei (Mill Valley: University Science Books)


\bibitem[]{}
Panagia, N. 1973, AJ, 78, 929

\bibitem[]{}
Petre, R., Szymkowiak, A. E., Seward, F. D., \& Willingale, R. 1988, ApJ, 335, 215

\bibitem[]{}
Raymond, J. C. 1979, ApJS, 39, 1

\bibitem[]{}
Reach, W. T., \& Rho, J. 1996, A\&A, 315, L277

\bibitem[]{}
Reach, W. T., \& Rho, J. 1998, preprint (astro-ph/9804142)

\bibitem[]{}
Reynolds, S. P., \& Moffett, D. A. 1993, AJ, 105, 2226

\bibitem[]{}
Rho, J.-H., \& Petre, R. 1996, ApJ, 467, 698

\bibitem[]{}
Rho, J.-H., Petre, R., Schlegel, E. M., \& Hester, J. J. 1994, ApJ, 430, 757

\bibitem[]{}
Richter, M. J., Graham, J. R., Wright, G. S., Kelly, D. M., \& Lacy, J. H. 1995a, ApJ, 449, L83

\bibitem[]{}
Richter, M. J., Graham, J. R., \& Wright, G. S. 1995b, ApJ, 454, 277

\bibitem[]{}
Schaller, G., Schaerer, D., Meynet, G., \& Maeder, A. 1992, A\&AS, 96, 269

\bibitem[]{}
Seta, M., et al. 1998, \apj, in press

\bibitem[]{}
Shelton, R. L., Cox, D. P., Maciejewski, W., Smith, R.,
Plewa, T., Pawl, A., \& R\'o\.zycska, M. 1998, \apj, submitted (astro-ph/9806090)

\bibitem[]{}
Shelton, R. L., Smith, R. K., \& Cox, D. P. 1995, BAAS, 27, 1213

\bibitem[]{}
Shull, J. M. 1980, ApJ, 237, 769

\bibitem[]{}
Shull, J. M., \& McKee, C. F. 1979, \apj, 227, 131

\bibitem[]{}
Slavin, J. D., \& Cox, D. P. 1992, \apj, 392, 131

\bibitem[]{}
Sturner, S. J., Skibo, J. G., Dermer, C. D., \& Mattox, J. R. 1997,
ApJ, 490, 619

\bibitem[]{}
Tauber, J. A., Snell, R. L., Dickman, R. L., \& Ziurys, L. M. 1994, ApJ, 421, 570

\bibitem[]{}
Turner, B. E., Chan, K.-W., Green, S., \& Lubowich, D. A. 1992, ApJ, 399, 114

\bibitem[]{}
van Dishoeck, E. F., Jansen, D. J. \& Phillips, T. G. 1993, A\&A, 279, 541

%\bibitem[]{}
%Wang, Z., \& Scoville, N. Z. 1992, ApJ, 386, 158

\bibitem[]{}
Wang, Z. R., Asaoka, I., Hayakawa, S., \& Koyama, K. 1992, PASJ, 44,303

\bibitem[]{}
Wheeler, J. C., Mazurek, T. J., \& Sivaramakrishnan, A. 1980, ApJ, 237, 781

\bibitem[]{}
White, R. L., \& Long, K. S. 1991, ApJ, 373, 543

\bibitem[]{}
Williams, J. P., Blitz, L., \& Stark, A. A. 1995, ApJ, 451, 252

\bibitem[]{}
Williams, J. P., \& McKee, C. F. 1997, ApJ, 476, 144

\bibitem[]{}
Wolszczan, A., Cordes, J. M., \& Dewey, R. J. 1991, ApJ, 372, L99

\bibitem[]{}
Wootten, A. 1977, ApJ, 216, 440

\bibitem[]{}
Wootten, A. 1981, ApJ, 245, 105


%





\end{thebibliography}
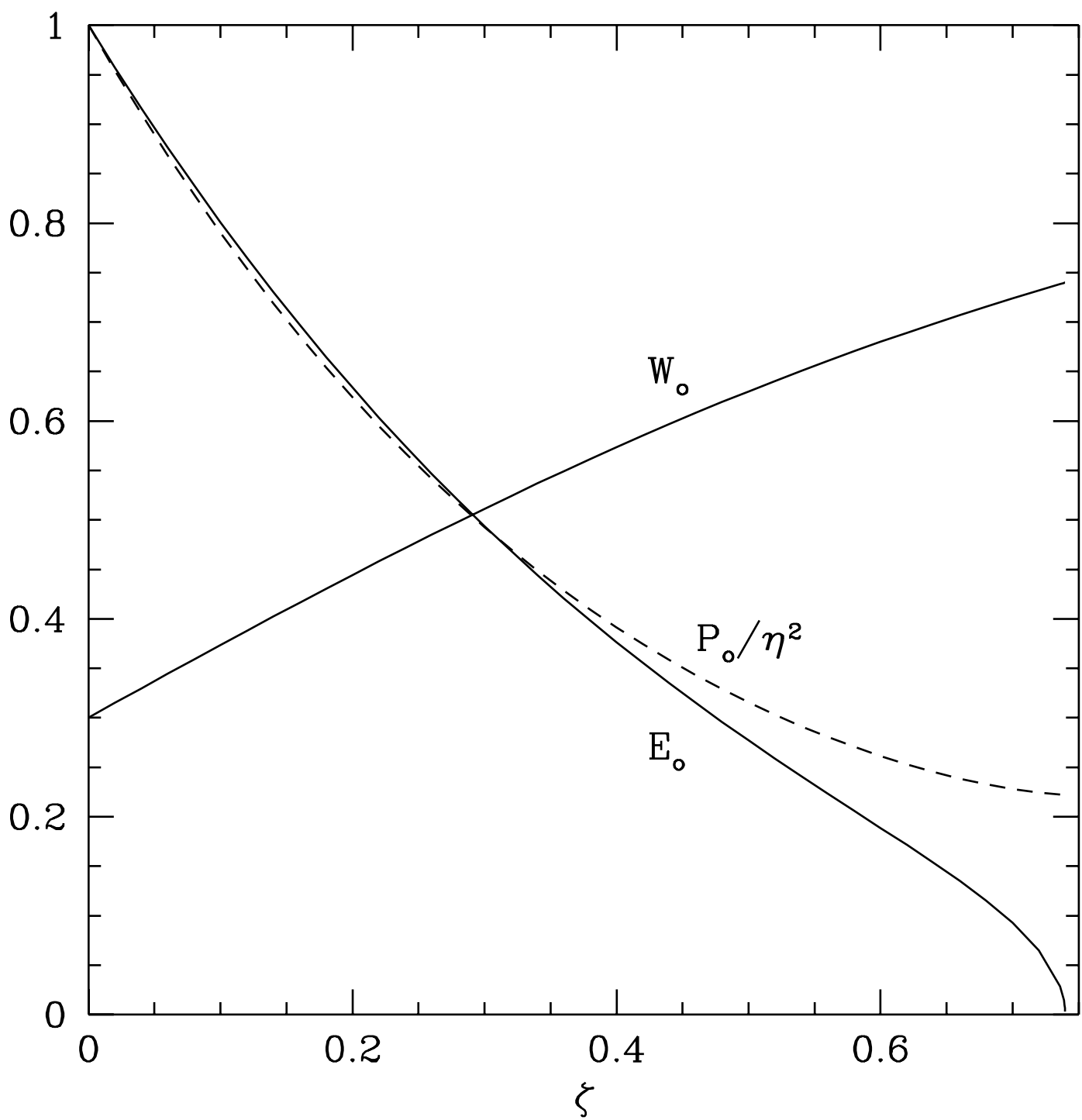
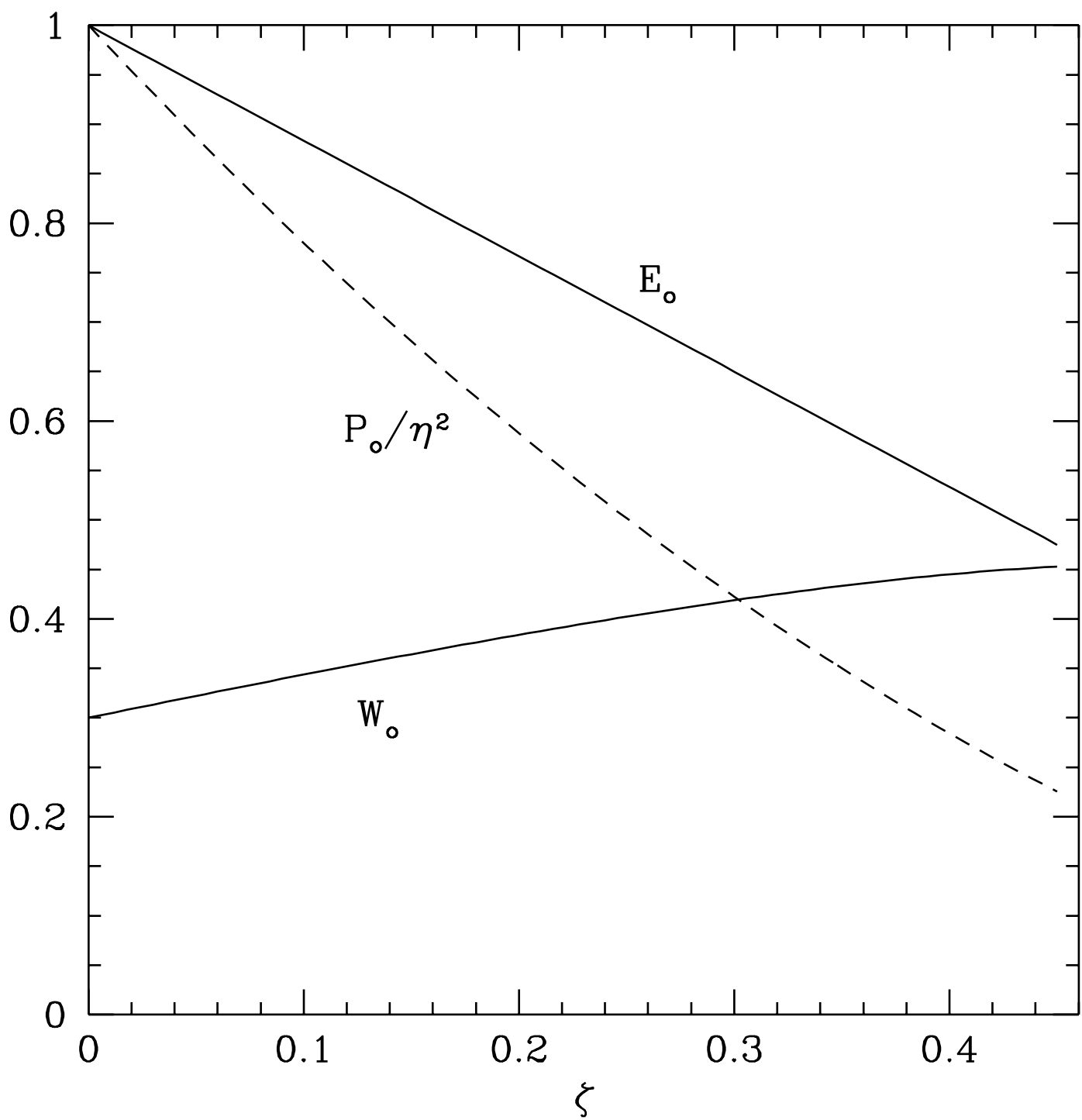
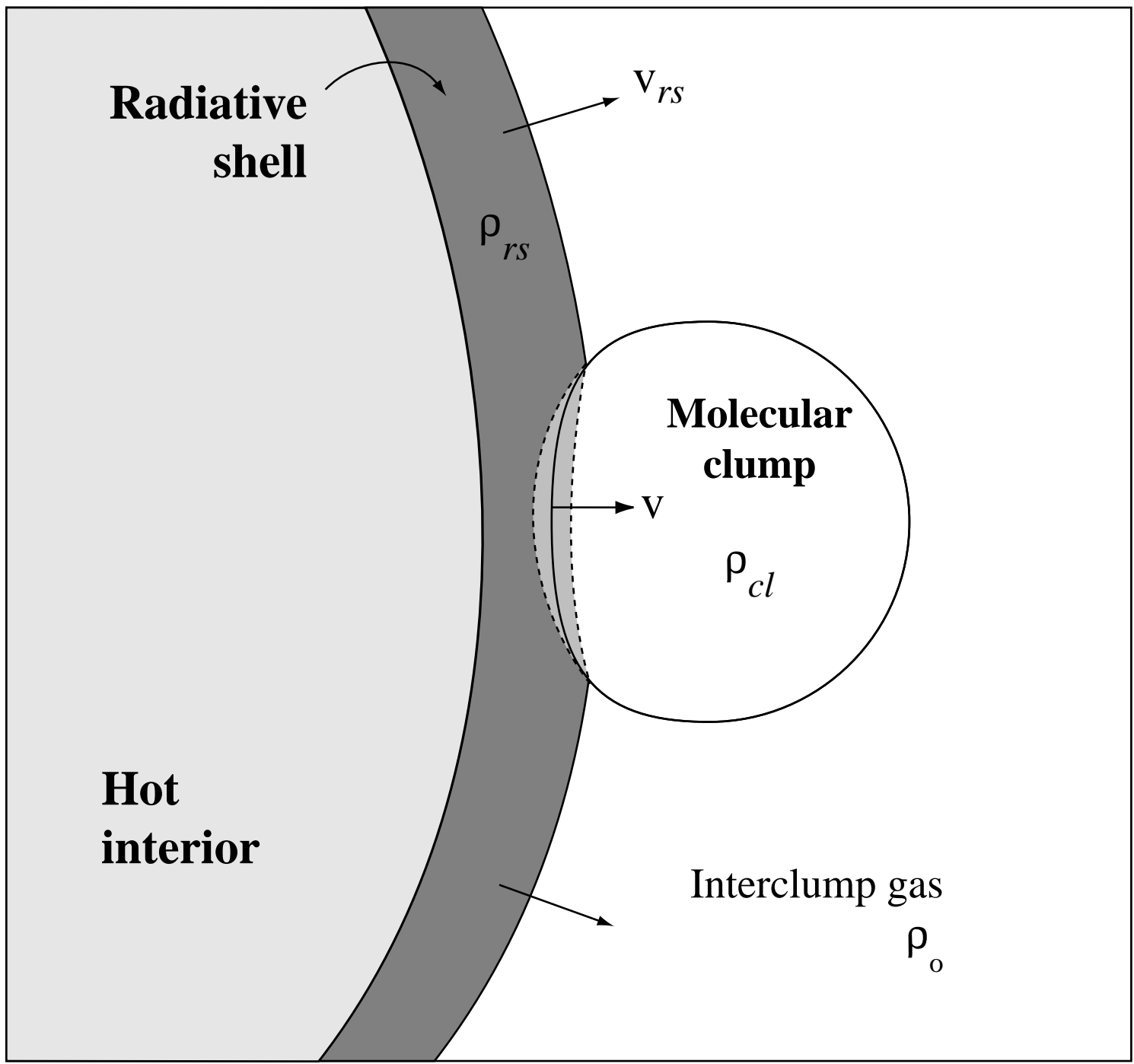
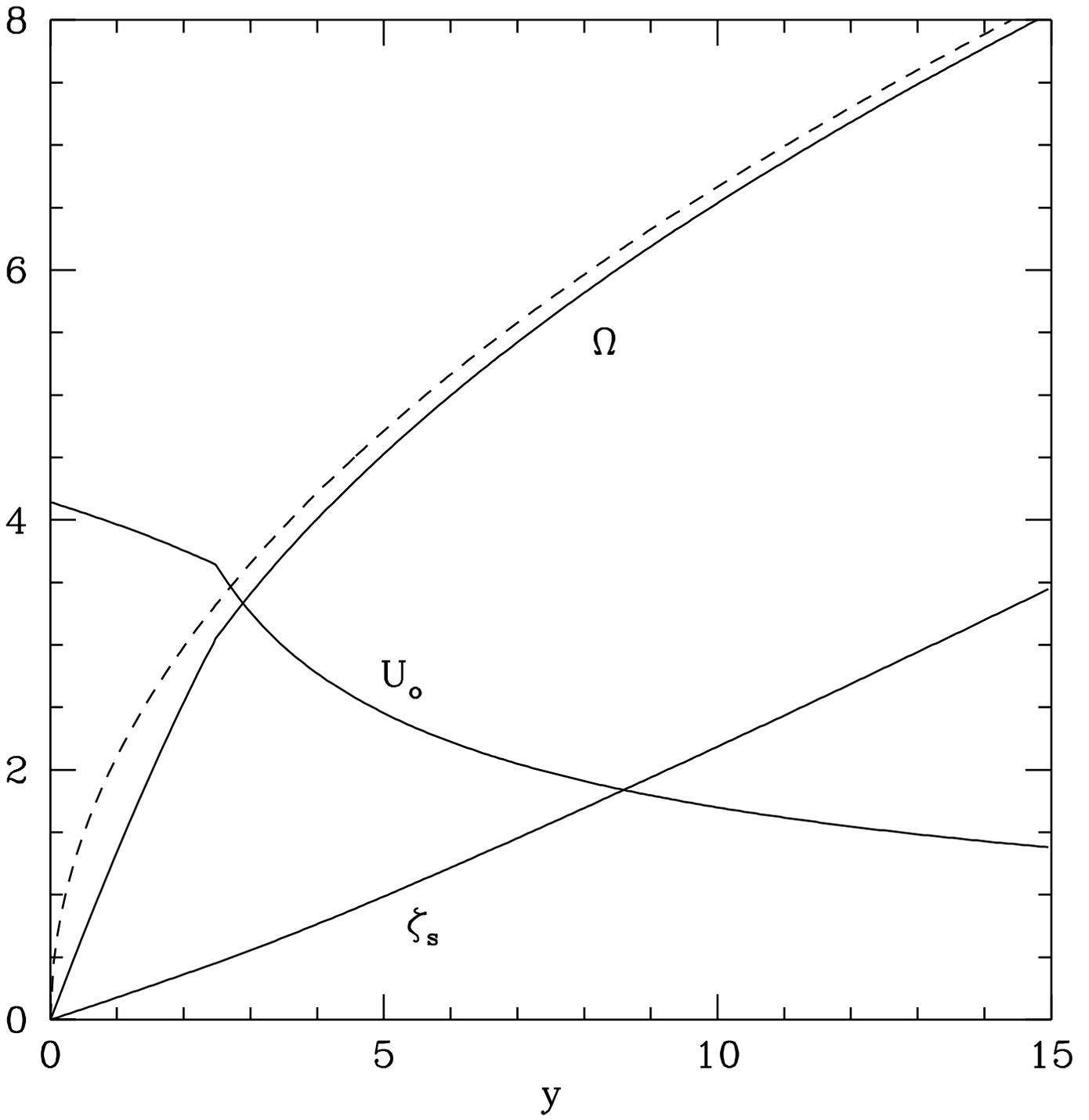
\end{document}